\documentclass[%
 reprint,
 compress,
 amsmath,amssymb,
 aps,
]{revtex4-1}
\usepackage{graphicx}% Include figure files
\usepackage{dcolumn}% Align table columns on decimal point
\usepackage{bm}% bold math
\usepackage{url}
\usepackage{float}
\usepackage{bm}
\usepackage{mathrsfs}
\usepackage{amssymb}
\usepackage{amsmath}
\usepackage{graphicx}
\usepackage{array}
\usepackage{epsfig,rotating,amsmath}
\usepackage{multirow}
\usepackage[table]{xcolor}
\usepackage{tabularx} 
\usepackage{color, colortbl}
\usepackage{graphicx}
\usepackage{subcaption}
\usepackage{adjustbox}
\usepackage{natbib}
\usepackage{pbox}
\usepackage{cases}
\usepackage{appendix}
\usepackage{etoolbox}
\usepackage[utf8x]{inputenc}
\usepackage{siunitx}

\definecolor{Gray}{gray}{0.9}
\begin{document}

%\preprint{APS/123-QED}

\title{Impact of the neutron-star deformability on equation of state parameters}% Force line breaks with \\

\author{C.Y. Tsang}
\author{M.B. Tsang}
\author{Pawel Danielewicz}
\author{W.G. Lynch}
\affiliation{%
 National Superconducting Cyclotron Laboratory and the Department of Physics and Astronomy \\
 Michigan State University, East Lansing, MI 48824 USA
}%

%\collaboration{HiRA Collaboration}%\noaffiliation

\author{F.J. Fattoyev}
\affiliation{
 Department of Physics, Manhattan College\\
 Riverdale, NY 10471, USA% with \\
}%

\date{\today}% It is always \today, today,

\begin{abstract}
We use a Bayesian inference analysis to explore the sensitivity of Taylor expansion parameters of the nuclear equation of state (EOS) to the neutron star dimensionless tidal deformability ($\Lambda$) on 1 to 2 solar masses neutron stars. A global power law dependence between tidal deformability and compactness parameter (M/R) is verified over this mass region. To avoid superfluous correlations between the expansion parameters, we use a correlation-free EOS model based on a recently published meta-modeling approach. We find that assumptions in the prior distribution strongly influence the constraints on $\Lambda$. The $\Lambda$ constraints obtained from the neutron star merger event GW170817 prefer low values of $L_\text{sym}$ and $K_\text{sym}$, for a canonical neutron star with 1.4 solar mass. For neutron star with mass $<1.6$ solar mass, $L_\text{sym}$ and $K_\text{sym}$ are highly correlated with the tidal deformability. For more massive neutron stars, the tidal deformability is more strongly correlated with higher order Taylor expansion parameters.

\end{abstract}

\pacs{Valid PACS appear here}% PACS, the Physics and Astronomy
\maketitle

%\tableofcontents

\section{\label{sec:level1}Introduction}

%A neutron star (NS) is the remnant of a supernova explosion of a massive star. Matter that makes up neutron star is predicted to be the densest material in the universe. Indeed, its density in the core is so high that it is energetically favorable for most protons and electrons to combine to form neutrons. Properties of neutron star capture the interest of nuclear physicists largely because it provides information on nuclear matter at high density.

%The equation that describes the properties of nuclear matter is called the Equation of State (EOS). In the past decade, various probes have been found to be sensitive to different density regions and constrain EOS at different densities. They combine to a compatible constraints up to three times the saturation density ($3\rho_0 = 3\times0.16\ \textrm{fm}^{-3}$). Together with astrophysical neutron star properties, a rough picture of the nuclear EOS has emerged. 

A neutron star (NS) is the remnant of a supernova explosion of a massive star. The interior of a NS contains the densest nuclear material in the universe. This matter is so dense that it becomes  energetically favorable for protons and electrons to combine and form neutrons. From densities ranging from somewhat below saturation density ($\rho_0 =\SI{0.155}{fm^{-1}}$ ) to $3\rho_0$, it is reasonable to describe NS matter as locally uniform nuclear matter composed mostly of neutrons. Study of NS is of great relevance to nuclear physics because of the information it can provide regarding the equation of state (EOS) of asymmetric nuclear matter at high density. Even though the current paper is self-contained with relevant materials detailed in appendix and extensive references, for those who are interested, Refs.~\cite{Li2019, Baldo2016, Lattimer2012, Holt2017}  provide more in depth discussions of the subjects.

Astrophysical NS properties, combined with constraints from nuclear observations, have provided a rough understanding of the EOS. Typical temperatures of NSs are low, $k_B T<\SI{1}{MeV}$; thus finite temperature effect is small and the main uncertainties in the EOS concern the relation between the pressure and energy density of nuclear matter at various baryon densities~\cite{Tsang2019533}. 

Measurements of collective flow and kaon production in energetic nucleus-nucleus collisions have constrained the EOS for symmetric matter, at densities up to 4.5$\rho_0$~\cite{Danielewicz2002,FUCHS2006, LYNCH2009}. Specifically, the symmetric matter constraints on pressure vs. density were determined in Ref.~\cite{Danielewicz2002} from the measurements of transverse and elliptical flow from Au+Au collisions over a range of incident energies from 0.3 to \SI{1.2}{GeV/u}. More recently, these constraints were confirmed in an independent analysis of elliptical flow data~\cite{LEFEVRE2016}. In Refs.~\cite{FUCHS2006, LYNCH2009}, a similar constraint from $1.2\rho_0$ to $2.2\rho_0$ was obtained from the Kaon measurements. These heavy ion constraints are consistent with the Bayesian analyses of the neutron-star mass-radius correlation in Ref.~\cite{Steiner2013}. 

Recent gravitational wave observations from LIGO collaboration~\cite{Abbott2017} opened a new window for understanding neutron-star matter. Specifically, the LIGO observation provides estimates for the tidal deformability, also known as tidal polarizability, a quantity that bears direct relevance to the nuclear EOS. 

The tidal deformability is induced when two NSs orbit around each other and tidal forces from each NS deforms its companion star. The mass quadrupole that developed in response to the external quadrupole gravitational field emerges as:
\begin{equation}
Q_{ij} = -\lambda E_{ij}.
\end{equation}
Here $E_{ij}$ is the external gravitational field strength and $\lambda$ is the tidal deformability. The orbital period of the inspiral differs from that of two point masses because the additional tidal deformation contributes to an overall orbital energy loss and changes the rotational phase. This difference is used to extract the dimensionless tidal deformability ($\Lambda$) of a NS~\cite{Damour1992, Flanagan2008}. Throughout this paper, tidal deformability given below always refers to the dimensionless tidal deformability,
%The mass quadrupole that developed in response to the external quadrupole gravitational field is related via:
%\begin{equation}
%Q_{ij} = -\lambda E_{ij}
%\end{equation}
\begin{equation}
\Lambda = \frac{\lambda c^{10}}{G^4M^5} = \frac{2}{3}k_2\Big(\frac{c^2R}{GM}\Big)^5,
\label{lambda}
\end{equation}
where $k_2$ is the second Love number~\cite{Damour2012, Binnington2009}. This whole expression, including the Love number, is sensitive to the nuclear EOS~\cite{Abbott2017, Postnikov2010, Piekarewicz2019}. Steps necessary to calculate $\Lambda$ for a given EOS are detailed in Appendix~\ref{TOVEquation}. Recent analysis of the gravitational wave data constrained this value to $\Lambda = 190^{+390}_{-120}$~\cite{Abbott2018}. 

Since most observables from nuclear structure experiments constrain the energy density and its derivatives near or somewhat below saturation density (See, for example, Refs. \cite{Kortelainen2010, Brown2013, Zhang2013, Danielewicz2016}), it is customary to approximate the EOS by a Taylor expansion about saturation density. We will explore the parameter space spanned by the derivatives of EOS with respect to density at $\rho_0$ and examine its correlation with $\Lambda$. 

Other studies have been carried out in placing tidal deformability constraints on these Taylor expansion parameters. They explored the constraints on different 2D parameter planes~\cite{Zhang2013, TSANG2019}, on a diverse set of models~\cite{Malik2018, Gil2019, Carson2019}, and with Bayesian analysis on EOSs from chiral effective field theory~\cite{Lim2018}. In this study, we will expand the analysis by employing a less restrictive form of EOS and exploring a larger parameter space by including higher order terms.

A family of theoretical EOS is needed to correlate the Taylor expansion parameters with the predicted $\Lambda$. One widely used family in astrophysics is the piece-wise polytropes~\cite{Lattimer2001}, but it is not suitable in this study because a Taylor expansion assumes that the EOS is analytic over the range of interest. As long as there is only one polytrope, a Taylor expansion is valid, but its validity does not extend past the point of connection between the original polytrope and the next. 

Another commonly used family is the Skyrme type EOS~\cite{Dutra2012}. It derives from simplified approximate nuclear interaction and relies on 15 free parameters in its expanded form. While it is shown to successfully reproduce various nuclear properties, it is difficult to explore new physics from the Taylor expansion parameters because they are strongly constrained by the form of the Skyrme interaction itself. It is difficult to access functional dependencies the Taylor expansion parameters that are not contained in the original choice for the Skyrme functional form~\cite{Khan2012, Khan2013}. 

In this study, an EOS from meta-modeling~\cite{Margueron2018} is used. By construction, their derivatives of different orders are independent of each other. 
This paper is organized as follows: In section \ref{Bay}, a brief description of Bayesian inference is provided. This is the statistical method employed in the extraction of EOS information from NS tidal deformability constraints. Section \ref{NEOS} describes our choice of EOS from meta-modeling approach in Ref.~\cite{Margueron2018} and how it is adopted to describe neutron star. In section \ref{Result1.4}, correlation between EOS parameters and tidal deformability of a 1.4 solar mass NS will be discussed. Section \ref{ResultHigher} extends the study to NSs of different masses and section \ref{Conclusion} summarizes our findings.

\section{\label{Bay}Bayesian inference}

We use Bayesian inference to study the influence of tidal deformability constraints from LIGO on nuclear-matter EOS parameters. These parameters are sampled with a prior probability distribution based on findings from literature and are then transformed into a distribution of neutron-star matter EOS. Through solving TOV equation, we are able to calculate the corresponding tidal deformabilities. By combining their prior distribution and likelihood, which indicates the compatibility between the calculated and the observed tidal deformability, Bayesian inference will assign probability for each EOS parameters with Bayes theorem:
\begin{equation}
P(\mathcal{M}) = \frac{1}{V_{\text{tot}}}w(\mathcal{M})p(\Lambda(\mathcal{M}))\prod_i g_i(m_i).
\label{bay}
\end{equation}
In this equation, $\mathcal{M}$ is the set of all EOS parameters, $m_i \in \mathcal{M}$ is one of the EOS parameters, $V_{\text{tot}}$ is the feature scaling constant, $p(\Lambda(\mathcal{M}))$ is the likelihood of a EOS calculated from its predicted $\Lambda$,  $g_i$ is the prior distribution of the $i^{th}$ parameter and $w(\mathcal{M})$ is the filter condition that filters out EOS parameter space that is nonphysical. 

The likelihood of EOS is the probability of having the observed LIGO event with the assumption that the given theoretical EOS is the ultimate true EOS. We will model the likelihood function as an asymmetric Gaussian distribution base on the extracted $\Lambda = 190^{+390}_{-120}$~\cite{Abbott2018} from GW170817.
\begin{equation}
p(\Lambda) = \begin{cases}
  \frac{1}{V}\exp(-\frac{(\Lambda - 190)^2}{2\times120^2}), & \text{if $\Lambda \leq 190$}\\
  \frac{1}{V}\exp(-\frac{(\Lambda - 190)^2}{2\times390^2}), & \text{if $\Lambda > 190$}.
\end{cases}
\end{equation}
In the above, $V$ is the feature scaling constant such that the likelihood function integrates to 1. 

The sought function is the probability distribution of EOS parameters rather than that for $\Lambda$, so prior distribution $g_i$ is required to convert between the two using Bayes theorem. A commonly used prior is the Gaussian distribution: 
\begin{equation}
g_i(m_i) = \frac{1}{\sqrt{2\pi\sigma_i^2}}\exp\Big(-\frac{(m_i - m_{i, \text{prior}})^2}{2\sigma_i^2}\Big),
\end{equation}
where $m_{i, \text{prior}}$ and $\sigma_i$ are the prior mean and standard deviation of the free parameters, respectively. They should be chosen to reflect our current understanding of those free parameters. 

Some parameter sets may yield nonphysical EOSs due to various additional considerations. The filter condition $w(\mathcal{M})$ takes that into account; it is set to 1 if the following three conditions of stability, causality and maximum mass, are all satisfied and it is set to 0, if not.

The stability condition rejects EOSs whose pressure decreases with energy density. 
%Such EOSs can only be accommodated if a phase change occurs. While it is not a required condition, but it contradicts our current assumption of the monotonically increasing pressure of nuclear matter.
Above the crust-core transition density, we require the EOSs to be mechanically stable with thermodynamical compressibility greater than zero, which means that the pressure of homogeneous matter does not decrease with density. For EOSs with negative compressibilities at density above the crust-core transition densities predicted by Eq. \eqref{transDens}, they will be rejected as being inconsistent with experimental information.

The requirement of causality rejects EOSs whose speed of sound is greater than the speed of light in the core region of their respective heaviest NS. The maximum mass condition rejects EOSs that fail to produce a NS of at least 2.04 solar mass in accordance with observation.~\cite{Demorest2010, Antoniadis2013}. 

Using the fact that the binary NS merger GW170817 detected by LIGO did not promptly produce a black hole, Ref.~\cite{Margalit2017} inferred that the heaviest possible NS should be around 2.17 solar mass. Other sources put the maximum mass at around 2.15-2.40 solar masses~\cite{Baym2019, Baoan2019, Shibata2017, Rezzolla2018, Ruiz2018, Zhou2018}. Neither of these constraints have been adopted in this work but can be implemented in the future.

The calculated probability distribution from Eq. \eqref{bay} is referred to as the posterior distribution. By comparing prior to posterior distribution, we will be able to infer the sensitivity of various EOS parameters to NS tidal deformability. By construction, priors of different free parameters in meta-modeling EOS are not correlated with each other, so any correlations in the posterior reflect the collective sensitivity of the Taylor expansion parameters to NS tidal deformability.

\section{\label{NEOS}Nuclear Equation of State}

\subsection{Parameters in nuclear EOS}

Nuclear matter is a theoretical construct composed of protons and neutrons. It resembles the core of ordinary nuclei where the neutron and proton densities are approximately uniform. Since the number of protons and neutrons are usually not far from each other in nuclei, we often expand EOS into the symmetric nuclear matter (SNM) term (isoscalar term) and a correction term for the deviation from SNM (isovector term), when proton densities and neutron densities are not identical as shown in Eq.~\eqref{EOSSplit} below. SNM refers to an infinite system where the density of proton equals to the density of neutrons. The EOS is commonly expanded as:
\begin{equation}
E(\rho, \delta) = E_{is}(\rho) + \delta^2E_{iv}(\rho).
\label{EOSSplit}
\end{equation}
In the above, $E_{is}$ is the isoscalar term, $E_{iv}$ is the isovector term, $\rho$ is the matter density and $\delta=(\rho_n - \rho_p)/(\rho_n+\rho_p)$ is called the asymmetry parameter. Nuclear structure probes are generally sensitive to the density region around saturation~\cite{Kortelainen2010, Brown2013, Zhang2013, Danielewicz2016} and as a result, derivatives of EOS with respect to density at this point are often used as empirical parameters to characterize the density and isospin dependence of the EOS. The derivatives are commonly expressed as  parameters in the Taylor expansion when EOS is expanded in terms of $x=(\rho - \rho_0)/(3\rho_0)$:

\begin{equation}
E_{is}(\rho) = E_0 + \frac{1}{2}K_{\text{sat}}x^2 + \frac{1}{3!}Q_{\text{sat}}x^3+\frac{1}{4!}Z_{\text{sat}}x^4+...
\label{ISExpansion}
\end{equation}
\begin{equation}
\begin{split}
E_{iv}(\rho) = &S_0 + Lx + \frac{1}{2}K_{\text{sym}}x^2 + \frac{1}{3!}Q_{\text{sym}}x^3 \\
&+\frac{1}{4!}Z_{\text{sym}}x^4+...
\end{split}
\label{IVExpansion}
\end{equation}
One focus of this paper is to explore the sensitivity between $\Lambda$ and $S$, $L$, $K$, $Q$, $Z$. Some families of EOS depend on density and asymmetry in a way that cannot be separated explicitly into the sum of two terms, but the isoscalar term is always well-defined:
\begin{equation}
E_{is}(\rho) = E(\rho, \delta=0).
\end{equation}
The isovector term can be defined as the second order Taylor expansion coefficient in $\delta$ around $\delta=0$ (not to be confused with Taylor EOS expansion  parameters which expands in $x$), 
\begin{equation}
E_{iv}(\rho) = \frac{1}{2}\frac{\partial^2 E(\rho, \delta)}{\partial\delta^2}\Big|_{\delta=0}.
\end{equation}
Likewise Taylor EOS  parameters can always be extracted from any nuclear EOS. This allows for comparison of variables across families of EOS. 

Another important quantity that characterizes nuclear matter properties is the effective mass $m^*(\rho, \delta)$. It is used to characterize the momentum dependence of nuclear interaction and it can be different for protons $m^*_\textrm{p}(\rho, \delta)$ and neutrons $m^*_\textrm{n}(\rho, \delta)$ depending on the condition which the nuclear matter is subjected to. It is generally assumed that $m^*_\textrm{p} = m^*_\textrm{n}$ in SNM. 

Comparison of effective mass is commonly carried out through the comparison of two quantities: the nuclear effective mass in SNM at saturation $m_\textrm{sat}^*$ and the splitting in neutron and proton effective masses in pure neutron matter (PNM) at saturation $\Delta m^* = m^*_\textrm{n} - m^*_\textrm{p}$. The choice of the two quantities mirrors the spirit of splitting EOS into isoscalar term and isovector term in Eq. \eqref{EOSSplit} in which contribution from SNM is separated from the correction factor that arises when matter is not symmetric.

Sometimes it is more convenient to express $m^*_\textrm{sat}$ and $\Delta m^*$ in terms of $\kappa_\textrm{sat}$, $\kappa_{sym}$ and $\kappa_{v}$:
\begin{equation}
\begin{split}
\kappa_{\text{sat}} &= \frac{m}{m^*_{\text{sat}}}-1=\kappa_s, \\
\kappa_{\text{sym}} &=\frac{1}{2}\Big(\frac{m}{m^*_n}-\frac{m}{m^*_p}\Big), \\
\kappa_{v} &= \kappa_\text{sat} - \kappa_\text{sym}.
\end{split}
\label{kappaform}
\end{equation}

The parameter $\kappa_{v}$ plays the role of the enhancement factor in Thomas-Reiche-Khun sum rule and it depends on the energy region of the resonance energy~\cite{LIPPARINI1989}. In this analysis, the effective masses will be expressed in terms of $m_\text{sat}^*/m$ and $\kappa_{v}$.

\subsection{EOS from a metamodeling approach}

Our studies utilizing the metamodeling analysis follow the approach of Ref.~\cite{Margueron2018}. Such metamodels for the EOS can be easily constructed with only Taylor expansion parameters and effective masses. The metalmodel EOS resembles Skyrme EOS with the same corresponding Taylor parameters to a greater extent than a simple power law expansion.

Four different \textit{empirical local density functionals} (ELF) meta-models are proposed in Ref.~\cite{Margueron2018}: ELFa, ELFb, ELFc and ELFd. ELFa does not produce vanishing energy as density approaches zero. ELFb does not converge to a typical Skyrme EOS even when identical Taylor parameters are used. ELFc does not have the shortcomings of EFLa and ELFb and closely resembles Skyrmes with similar Taylor parameters. Although ELFd agrees with Skyrmes  better, it relies on high density information that is not well constrained by experiments. 

From the above considerations, we adopt ELFc in this study. Similar choice is also made in other recent studies~\cite{Tews2018_2, Guven2020}. The formulation of ELFc is detailed in Appendix \ref{mm_mapping}. As assessed in Ref.~\cite{Margueron2018}, the following choices of parameters have been accurately constrained by nuclear experiment and are fixed in the analysis: $E_{\text{sat}}=\SI{-15.8}{MeV}$, $\rho_0=\SI{0.155}{fm^{-3}}$.

\subsection{Thermodynamic relations}

Additional characteristics of nuclear matter can be inferred using thermodynamic equations once an EOS is specified. The pressure at various densities $P(\rho)$ is related to the derivative of the energy:
\begin{equation}
P(\rho) = \rho^2\frac{\partial E(\rho, \delta)}{\partial \rho}.
\end{equation}
The adiabatic speed of sound can then be calculated~\cite{Fluid}:
\begin{equation}
\Big(\frac{v_s}{c}\Big)^2 = \Big(\frac{\partial P}{\partial \mathcal{E}}\Big)_S,
\end{equation}
where $\mathcal{E} = \rho (E + mc^2)$ is the energy density of the material including mass density. This implies any thermodynamic stable EOS must satisfy $\Big(\frac{\partial P}{\partial \mathcal{E}}\Big)_S >0$. Furthermore, since information cannot travel faster than the speed of light due to causality, the inequality $v_s < c$ must hold for all densities relevant to NS. This may not be always true for ELFc. To stay physical, we will switch from ELFc to an expression for the stiffest possible EOS whenever causality is violated:
\begin{equation}
P_{\text{Stiffest}}(\mathcal{E}, v_s, \mathcal{E}_0, P_0) = \Big(\frac{v_s}{c}\Big)^2(\mathcal{E} - \mathcal{E}_0) + P_0.
\label{stiffestEOS}
\end{equation}
This equation represents a EOS with constant speed of sound $v_s$ and $v_s=c$ yields the stiffest possible EOS~\cite{Lattimer2015}. Here $\mathcal{E}_0$ and $P_0$ are reference values of energy density and pressure, respectively. The reference values can be adjusted to match the conditions at a specific density where energy density and pressure are known. The switch in EOS avoids superfluous rejection when causality is considered.
%, but those EOSs are stiffer than the true neutron-star matter.

%This equation will be used when exploring extremely high density region where a phase change to exotic matter is possible. We do not assume a phase change occurring in the NS. Instead we use the stiffest EOS in those regions to avoid superfluous rejection when causality is considered. 

\subsection{Structure of a NS and modifications on the nuclear EOS}

\begin{table*}[!ht]
\begin{center}
\caption{Summary information of various models~\cite{Margueron2018}. The bottom half shows characteristics of the prior and posterior distribution respectively.}\label{tab:OtherModelTable}
\small
\setlength{\tabcolsep}{0pt}
\begin{tabular}{lccccccccc}
\hline  
 &    \pbox{50cm}{$\ \ L_{\text{sym}}\ \ $\\$(MeV)$} & \pbox{20cm}{$\ \ K_{\text{sym}}\ \ $\\$(MeV)$} & \pbox{20cm}{$\  \ K_{\text{sat}}\ \ $\\$(MeV)$} & \pbox{20cm}{$\ \ Q_{\text{sym}}\ \ $\\$(MeV)$} & \pbox{20cm}{$\ \ Q_{\text{sat}}\ \ $\\$(MeV)$} & \pbox{20cm}{$\ \ Z_{\text{sym}}\ \ $\\$(MeV)$} & \pbox{20cm}{$\ \ Z_{\text{sat}}\ \ $\\$(MeV)$} & $\ \ \frac{m^{*}_{\text{sat}}}{m}\ \ $ & $\ \ \kappa_v\ \ $ \\[10pt]
 \hline
  \\
 \\[-1em]
 Skyrme Average & 49.6 & -132 & 237 & 370 & -349 & -2175 & 1448 & 0.77   & 0.44\\[2pt]
                                                                 
 Skyrme $\sigma$ & 21.6 & 89   & 27  & 188 & 89   & 1069  & 510  & 0.14   & 0.37 \\[2pt]
                                                                 
 RMF Average & 90.2    & -5   & 268 & 271 & -2   & -3672 & 5058 & 0.67   & 0.40 \\[2pt]
                                                                 
 RMF $\sigma$ & 29.6    & 88   & 34  & 357 & 393  & 1582  & 2294 & 0.02   & 0.06 \\[2pt]
                                                                 
 RHF Average & 90.0    & 128  & 248 & 523 & 389  & -9956 & 5269 & 0.74   & 0.34 \\[2pt]
                                                                 
 RHF $\sigma$ & 11.1    & 51   & 12  & 237 & 350  & 4156  & 838  & 0.03   & 0.07 \\[2pt]
 \hline      

 %\rowcolor{Gray}                                            
 Weighted Average & 69.0 & -45.3& 248 & 367 & -114 & -3990 & 3310 & 0.712  & 0.42 \\[2pt]
 %\rowcolor{Gray}        
 Weighted $\sigma$ & 20.1& 70.8 & 18.3& 214 & 200  & 1530  & 989  & 0.06 & 0.17 \\[2pt]
 \hline
 %\rowcolor{Gray}        
 Posterior Average & 71.6 & -76.9 & 245 & 436 & -97 & -3410 & 3490 & 0.74 & 0.41 \\[2pt]
 %\rowcolor{Gray}        
 Posterior $\sigma$ & 16.5 & 66.0 & 23 & 219 & 202 & 1710 & 970	& 0.07 & 0.25 \\[2pt]
 \hline

\end{tabular}
\end{center}
\end{table*}

Neutron stars are more than a “giant nucleus” described in Ref.~\cite{Yakovlev2013}. There are structural changes at various density regions as a result of a competition between the nuclear attraction and the Coulomb repulsion. The dynamics of the outermost layers of NSs is described mostly by the Coulomb repulsion and nuclear masses, where nuclei arrange themselves in a crystalline lattice. As the density increases, it becomes energetically favorable for the electrons to capture protons, and the nuclear system evolves into a Coulomb lattice of progressively more exotic, neutron-rich nuclei that are embedded in a uniform electron gas. This outer crustal region exists as a solid layer of about 1 km in thickness~\cite{Piekarewicz2019}. 

At intermediate densities of sub-saturation, the spherical nuclei that form the crystalline lattice start to deform to reduce the Coulomb repulsion. As a result, the system exhibits rich and complex structures that emerge from a dynamical competition between the short-range nuclear attraction and the long-range Coulomb repulsion~\cite{NeutronStarsI}. 

At densities of about half of the nuclear saturation, the uniformity in the system is restored and matter behaves as a uniform Fermi liquid of nucleons and leptons. The transition region from the highly ordered crystal to the uniform liquid core is very complex and not well understood. At these regions of the inner crust which extend about 100 meters, various topological structures are thought to emerge that are collectively referred to as “nuclear pasta”. Despite the undeniable progress~\cite{pasta1, pasta2, pasta3, pasta4, pasta5, pasta6, pasta7, pasta8, pasta9, pasta10, pasta11, pasta12, pasta13, pasta14, pasta15, pasta16, pasta17, pasta18,  Horowitz2004,Horowitz2004_2,Horowitz2005,Watanabe2003,Watanabe2005,Watanabe2009,Schneider2013,Horowitz2015,Caplan2015,Magierski2002,Chamel2005,Newton2009,Schuetrumpf2015,Fattoyev2017} in understanding the nuclear-pasta phase since their initial prediction over several decades ago~\cite{Ravenhall1983, Hashimoto:1984, Oyamatsu:1984}, there is no known theoretical framework that simultaneously incorporates both quantum-mechanical effects and dynamical correlations beyond the mean-field level. As a result, a reliable EOS for the inner crust is still missing. %Fortunately, the contributions of the pasta region to the tidal deformability are not that significant because the pressure is dominated by the electron degeneracy pressure for the crust except in the comparatively thin pasta region.

The matter in the core region of NS can be described as uniform nuclear matter where neutron, proton, electrons and muons exist in beta equilibrium~\cite{NeutronStarsI}. Although a phase change and exotic matter such as hyperons~\cite{Ambartsumyan1960,Chatterjee2016,NeutronStarsI} could appear in the inner core region, there is currently no direct evidence of their existence. In this work, we calculate the EOS in this region by assuming that the neutron-star matter is composed of nucleons and leptons only.

Due to the rich structure of NS, the nuclear EOS needs to be contextualized before it can be used for NS properties calculation. To begin with, crustal EOS should be used at density below transition density $\rho_T$. Normally the determination of $\rho_T$ requires complicated thermodynamic calculations, but some simple relationship has been found between transition densities and Taylor parameters of the EOS in Ref.~\cite{Ducoin2011} that greatly simplifies its calculation. In this study, the following equation is used to determine $\rho_T$:
\begin{equation}
\rho_T = (-3.75\times10^{-4} L_{\text{sym}} + 0.0963)\ \text{fm}^{-3}.
\label{transDens}
\end{equation}
Outer and inner crust exhibit different physical properties and should be described by different EOSs. For the outer crust, EOS provided by Ref.~\cite{Baym1971} is used in this analysis. For the inner crust, spline interpolation the region of $0.3\rho_T < \rho < \rho_T$ is reserved for a smooth transition between the outer crust and outer core.
%The lack of adjustable parameters in crustal EOS means that this EOS cannot connect smoothly to ELFc at the transition density. 
While this connection region cannot precisely describe crustal dynamics, tidal deformability does not appear to be sensitive to the choice of the crustal details for NS~\cite{Piekarewicz2019,Ji2019,Perot2020}.

The outer core region $\rho > \rho_T$ is characterized by the EOS of a beta equilibrated system of protons, neutrons, electrons and muons. Proton and neutrons are collectively described by EFLc while electrons and muons are modeled as relativistic Fermi gases. Equilibrium is attained by minimizing the Helmholtz free energy at different densities. If the speed of sound for EOS reaches the speed of light at density $\rho = \rho_c$, it will switch to the stiffest possible EOS of Eq. \eqref{stiffestEOS} at higher densities to comply with causality condition. If ELFc does not violate causality at all densities relevant to NSs, then $\rho_c=\infty$ and Eq.~\eqref{stiffestEOS} is never used.

To summarize, EOS of the neutron-star matter is formulated as follows:

\begin{equation}
P(\mathcal{E}) = \begin{cases}
  P_{\text{crust}}(\mathcal{E}), & \text{if $0 < \rho < 0.3\rho_T$}\\
  P_{\text{spline}}(\mathcal{E}), & \text{if $0.3\rho_T < \rho < \rho_T$}\\
  P_{\text{EFLc}}(\mathcal{E}), & \text{if $\rho_T < \rho < \rho_c$}\\
  P_{\text{stiffest}}(\mathcal{E}, c, \mathcal{E}_0, P_0) & \text{if $\rho_c < \rho$}.
\end{cases}
\end{equation}
In the above equation, $\mathcal{E}_0$ and $P_0$ are the energy density and pressure from beta-equilibrated EFLc at $\rho_c$ respectively, $P_{\text{crust}}$ is the pressure from crustal EOS and $P_{\text{ELFc}}$ is the pressure from beta-equilibrated ELFc EOS. $P_{\text{spline}}$ and $\mathcal{E}_{\text{crust}}$ govern the cubic spline that smoothly connects $P_{\text{crust}}$ to $P_{\text{ELFc}}$ and $\mathcal{E}_{\text{crust}}$ to $\mathcal{E}_{\text{EFLc}}$ respectively.

\section{\label{Result1.4}Results for a 1.4-solar mass NS}

\begin{figure*}[!ht]
\includegraphics[width=0.9\linewidth]{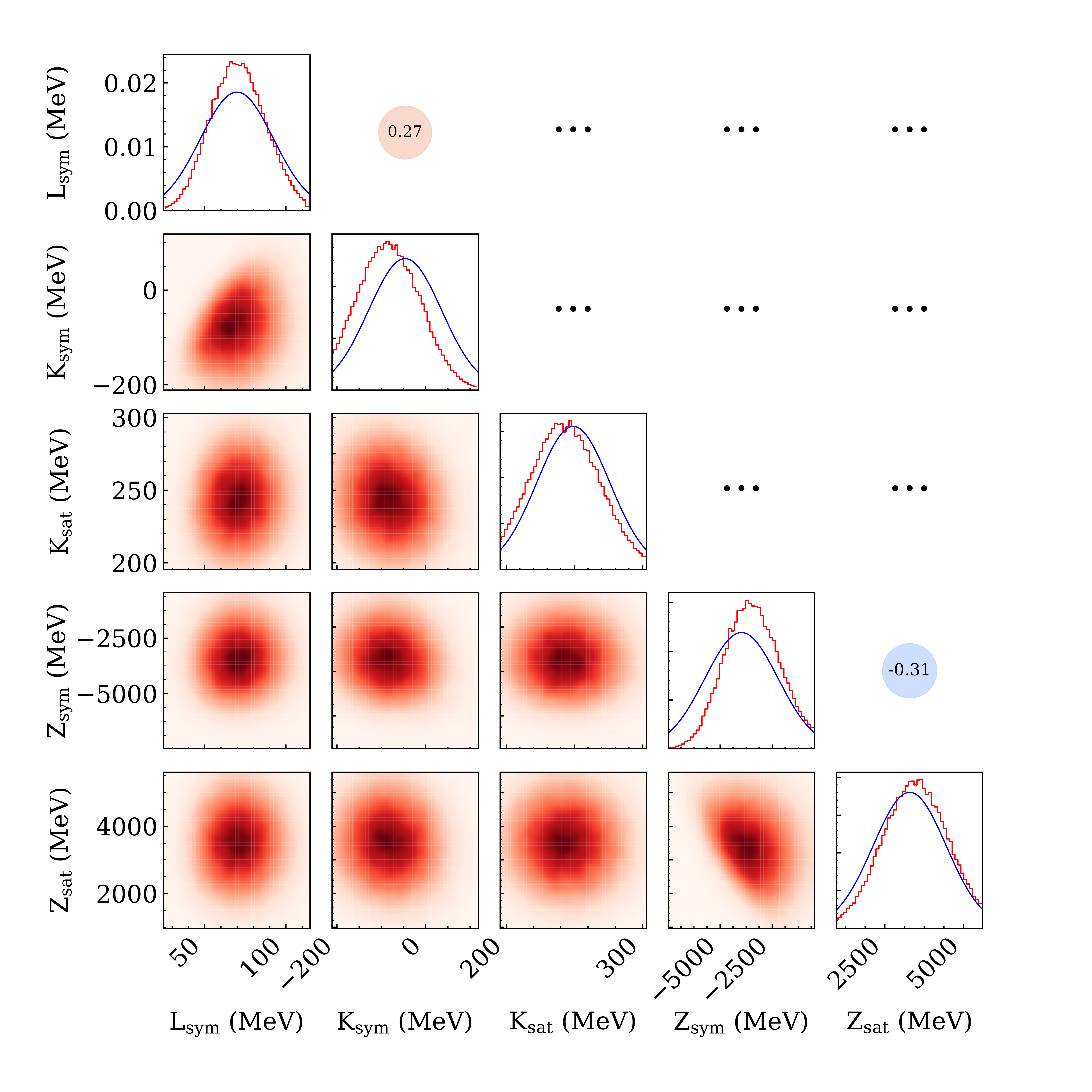}
\captionof{figure}{Bivariate characteristics of posterior likelihood distributions. Three regions can be distinguished. The lower triangle panels show likelihood distributions, with intensity proportional to distribution value, for pairs of Taylor parameters. The diagonal panels display prior (blue) and marginalized posterior (red) distributions for each parameter. The upper triangular region shows Pearson correlation coefficient for parameter pairs. Three dots indicate weak correlations with magnitude less than 0.1. }
\label{Correlation}
\end{figure*}

A total of 1,500,000 EOSs have been sampled and 682,652 of them satisfy all of our constraints. Only 11,711 EOSs apply to all densities without switching to the stiffest EOS. 

Prior distributions of the parameters should reflect our initial belief of those quantities before information on tidal deformability is taken into account. For this, we rely on Ref.~\cite{Margueron2018} which summarizes the distributions of EOS parameters from three phenomenological families, Skyrme, relativistic mean field (RMF) and relativistic Hartee-Fock (RHF). The mean and standard deviation of the parameters for each family are tabulated in the first six rows of Table \ref{tab:OtherModelTable}. In this study, the prior means and standard deviations are the weighted average values of the 3 families, with weights of 0.500,\ 0.333,\ 0.167 respectively. The weights reflect our confidence in of the models. We give Skyrme EOS the most weight as it is the most heavily employed parametrization in a myriad of nuclear predictions~\cite{Dutra2012}. These relative weights are ad hoc, but should cover most plausible parameter spaces. Prior means and standard deviations are listed in the seventh and eighth row in table \ref{tab:OtherModelTable} respectively.

The posterior distributions of Taylor expansion parameters are represented in Fig.~\ref{Correlation}. The lower triangular plots show the bivariate distributions for two parameters. The diagonal plots show the prior (blue curves) and marginalized posterior distributions  (red curves) for individual parameters. The upper triangle displays the Pearson correlation coefficients for parameter pairs:
\begin{equation}
\rho_{X,Y}=\frac{\mathbf{E}[(X-\bar{X})(Y-\bar{Y})]}{\sigma_X\sigma_Y},
\end{equation}
where $\mathbf{E}$ is the expectation value and $\sigma_{X}$ and $\sigma_{Y}$ are the standard deviations of the parameters distributions. The Pearson coefficient ranges from -1 to 1 and its absolute value reflects the strength of the correlation. A positive value close to 1 indicates a strong correlation and a negative value close to -1 indicates strong anti-correlation while a value close to 0 indicates lack of correlation~\cite{PC}. Only bivariate distributions between $L_{\text{sym}}$, $K_{\text{sym}}$, $K_{\text{sat}}$, $Z_{\text{sym}}$ and $Z_{\text{sat}}$ are shown because the higher order parameters do not seem to be influenced by our tidal deformability constraints. The full correlation plot is included in Appendix \ref{secfullcor}. Characteristics of the probability distribution are summarized in the bottom two rows of Table. \ref{tab:OtherModelTable}.

\begin{figure}
\includegraphics[width=1\linewidth]{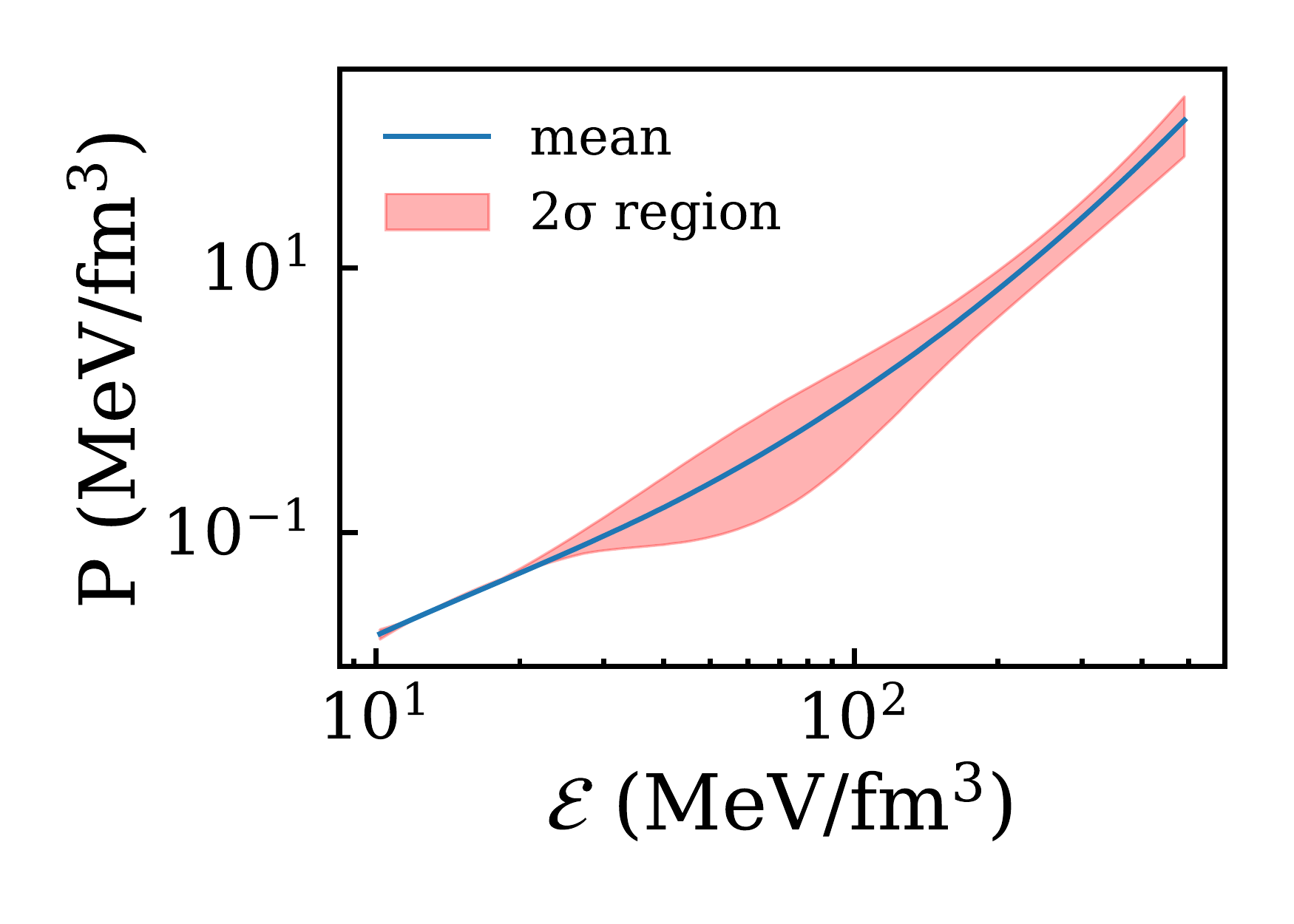}
\caption{Distribution of EOSs sampled from the posterior. The divergence above energy density $\sim > \SI{20}{MeV/fm\cubic}$ coincides with the transition from outer crust to spline connection.}
\label{GoodExample}
\end{figure}

Fig.~\ref{GoodExample} shows the mean and 2$\sigma$ region spanned by the EOS in the posterior. The 2$\sigma$ region converge to a line for $\mathcal{E} \lesssim \SI{20}{MeV/fm\cubic}$, which corresponds to the outer crust. Since we connect all EOSs to the crustal EOS given by Ref.~\cite{Baym1971}, this convergence is expected. From around \SI{20}{MeV/fm\cubic} to \SI{70}{MeV/fm\cubic}, the spline connection kicks in and manifests in the broadening of pressure.

\begin{figure}[!ht]
\includegraphics[width=1\linewidth]{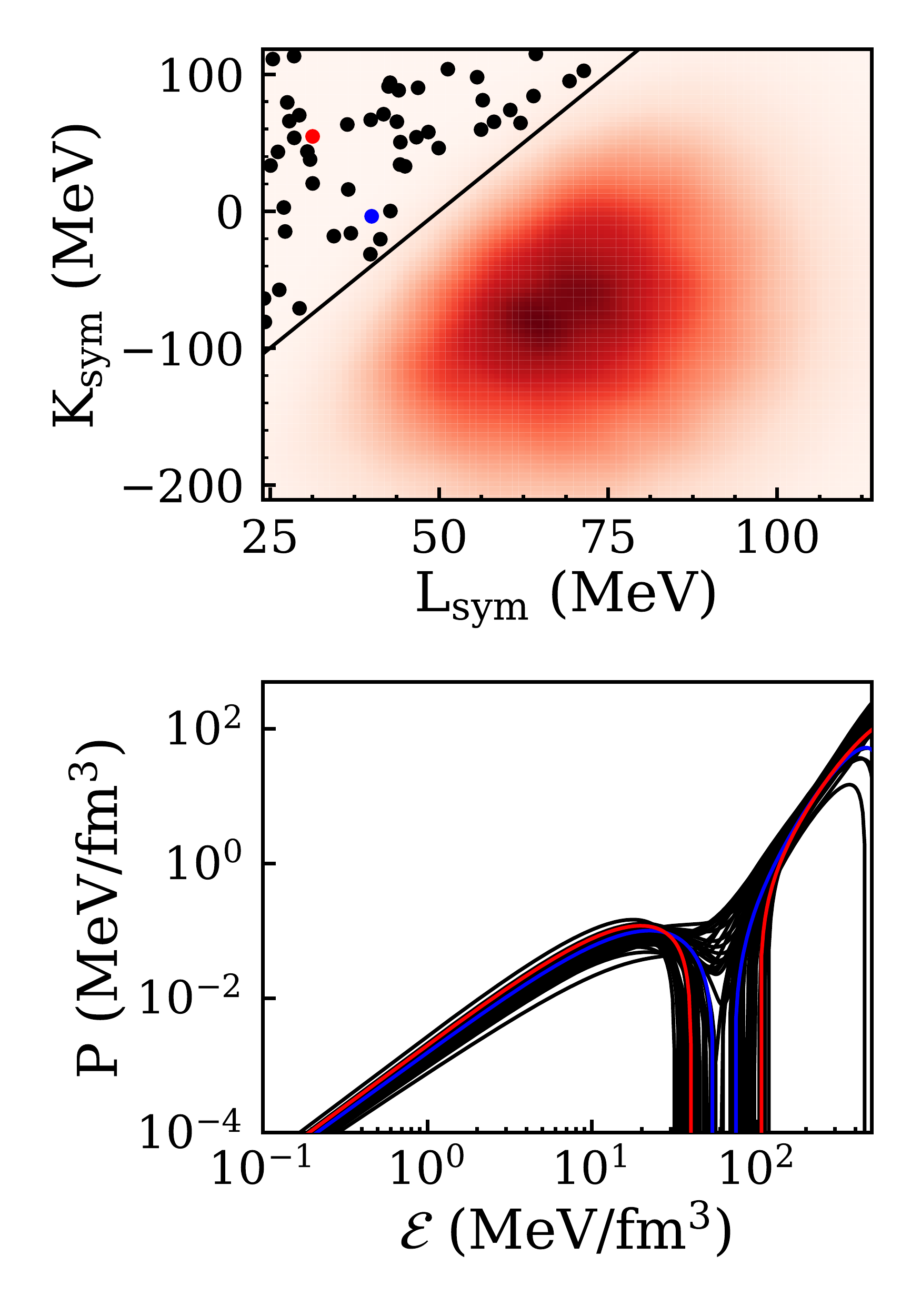}
\caption{(Upper panel) The 50 dots in the upper left hand corner of $K_\text{sym}$ vs. $L_\text{sym}$ correspond to 50 randomly chosen parameter space within the stability cut-off region. (Lower panel) Unstable EOS corresponding to the 50 dots. The red and blue line correspond to the red and blue point in the upper panel respectively. They are highlighted to showcase how a typical EOS in the cut-off region looks like. }
\label{BadExample}
\end{figure} 

The cut-offs in the lower left corner of $Z_{\text{sym}}$ vs. $Z_{\text{sat}}$ distribution and the upper left corner of $K_{\text{sym}}$ vs. $L_{\text{sym}}$ distribution in Fig.~\ref{Correlation} are the consequence of stability condition. At such extreme values, speed of sound may be imaginary when extrapolating to NS of 2 solar masses. This is evident in Fig.~\ref{BadExample} in which 50 randomly selected EOSs from the cut-off region in $K_\text{sym}$ vs. $L_\text{sym}$ are shown in the lower panel. The pressure for those EOSs do not increase monotonically with the energy density and become mechanically unstable. These EOSs are discarded. 

The posterior distributions of $K_{\text{sym}}$ and $Z_{\text{sym}}$ differ from the prior distributions significantly. The tidal deformability constraint favors lower $K_{\text{sym}}$ region. The inference also narrows the range of possible $L_{\text{sym}}$. Parameters such as $K_\text{sat}$ and $Z_\text{sat}$, whose posterior distributions are not altered significantly reflect that they are not sensitive to the tidal deformability constraints. 

%While this Bayesian analysis is well suited to discuss the sensitivity of the deformability to the Taylor expansions parameters $L_\text{sym}$, $K_\text{sym}$, $K_\text{sat}$, etc., it has some limitations. In particular, we note that the most probable value for $\Lambda$ is $624\pm129$. This value exceeds the value obtained by Ref.~\cite{Abbott2018} from the analysis of the GW170817 due to the strong bias of the prior distributions of RMF and RHF interactions. The latter prefer a most probable value for the deformabilitity of $\Lambda\sim750$ while the distribution of Skyrme interactions prefers a lower deformabiltiy of $\Lambda\sim400$. The raises the question of the dependence of the GW170817 results on the prior distributions of the EOS used in the analysis to extract the deformability~\cite{Abbott2018}.

While this Bayesian analysis is well suited to discuss the sensitivity of the deformability to the Taylor expansions parameters $L_\text{sym}$, $K_\text{sym}$, $K_\text{sat}$, etc., it has some limitations. In particular, we note that the prior and post distributions of $\Lambda$ as shown in Fig.~\ref{FullCor} (row 2 column 10 in Appendix~\ref{secfullcor}) are drastically different, probably as a consequence of the narrow prior distributions of the Taylor expansion parameters listed in Table~\ref{tab:OtherModelTable}. This reflects the strong sensitivity of $\Lambda$ to the prior distributions of the EOS. Furthermore, the posterior distribution of $\Lambda$ is much sharper and peaked at $624\pm129$ which exceeds the value of $190^{+390}_{-120}$ obtained in Ref.~\cite{Abbott2018} from the analysis of the GW170817. While the GW constraint reflects the high density of NS core, the prior distributions of the Taylor expansion parameters do not have rigorous laboratory constraints at high density region where $\Lambda$ is determined.

\section{\label{ResultHigher}Neutron star with different masses}

\begin{table}%[htbp]
\begin{center}

\small
\setlength\tabcolsep{4pt}
\caption{Predicted tidal deformability for NS of different masses}
\label{tab:PosteriorTable}
\begin{tabular}{lrrrrr}
\hline 
 & \pbox{20cm}{$\Lambda(1.2)$} & \pbox{20cm}{$\Lambda(1.4)$} & \pbox{20cm}{$\Lambda(1.6)$} & \pbox{20cm}{$\Lambda(1.8)$} & \pbox{20cm}{$\Lambda(2.0)$}\\ [10pt]
 \hline
 \\[-1em]
 Posterior Average & 1490 & 624 & 281 & 132 & 64 \\[2pt]
 Posterior $\sigma$ & 310 & 129 & 61 & 31 & 17\\[2pt]
 \hline
\end{tabular}
\end{center}
\end{table}

\begin{figure*}
\includegraphics[width=1\linewidth]{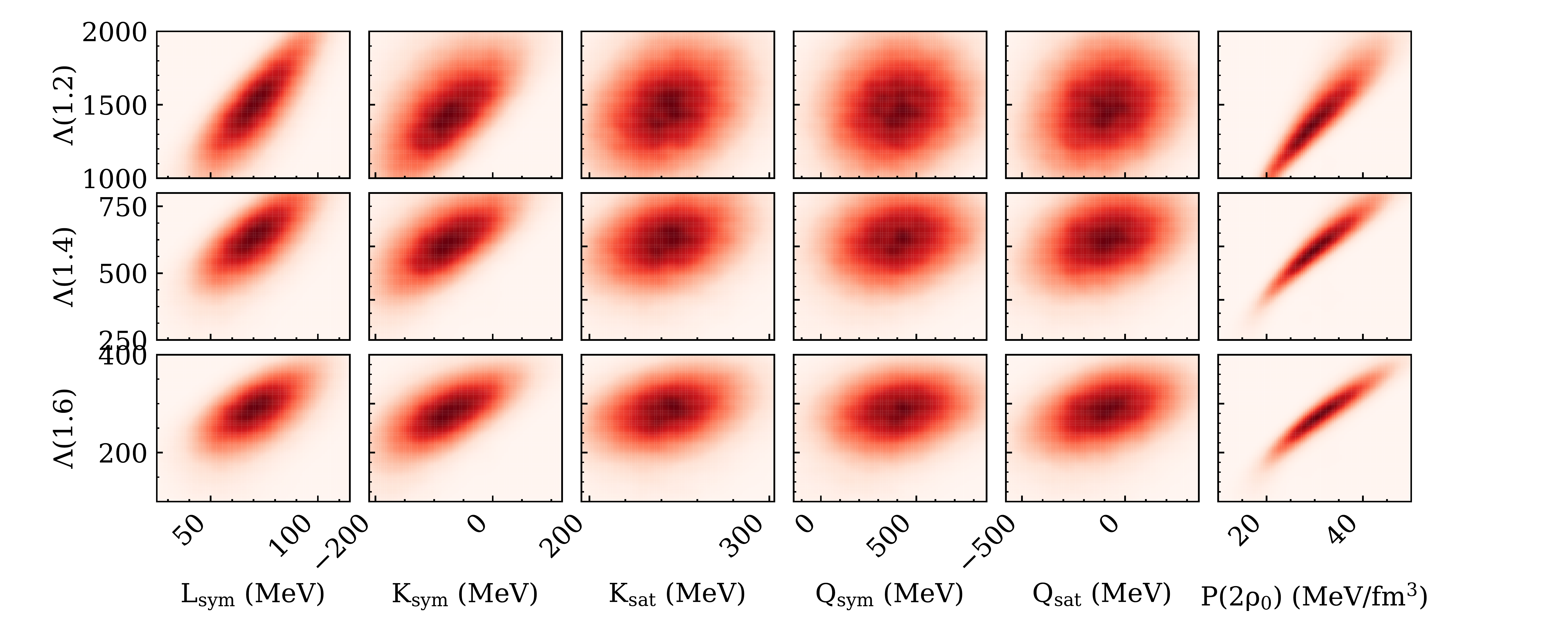}
\caption{Bivariate distributions between deformabilities with NS of different masses and Taylor parameters. Correlation with tidal deformability is clearly seen with $L_{\text{sym}}$, $K_{\text{sym}}$ and $P(2\rho_0)$.}
\label{Cor}
\end{figure*}

While the chirp mass of GW170817 has been determined quite accurately~\cite{Abbott2017}, the exact masses of the two neutron stars or their mass ratios are not known~\cite{Abbott2017}. In anticipation that more merger events involving different NS masses than the nominal NS mass of 1.4 solar mass are observed in the future~\cite{Abbott2020}, we use the posterior EOS distributions to predict the deformability of NS with different masses
. The posterior EOS distributions can be used to predict the deformability of NS with different masses. In Table \ref{tab:PosteriorTable}, we provide our predictions for the tidal deformabilities for NS with
 1.2, 1.4, 1.6, 1.8 and 2 solar mass using this group of EOSs weighted by their posterior distributions. To show the sensitivity of these predictions to the Taylor parameters, Fig.~\ref{Cor} shows the bivariate distributions between the Taylor parameters of the posterior distributions and the predicted tidal deformabilities of different stellar masses. We find that $\Lambda$ is more strongly correlated with $L_\text{sym}$  and $K_\text{sym}$  than it is with higher order Taylor expansion parameters. The sensitivity to $K_\text{sym}$ increases, while the sensitivity to $L_\text{sym}$ decreases, with stellar mass. 

\begin{figure}
\includegraphics[width=1\linewidth]{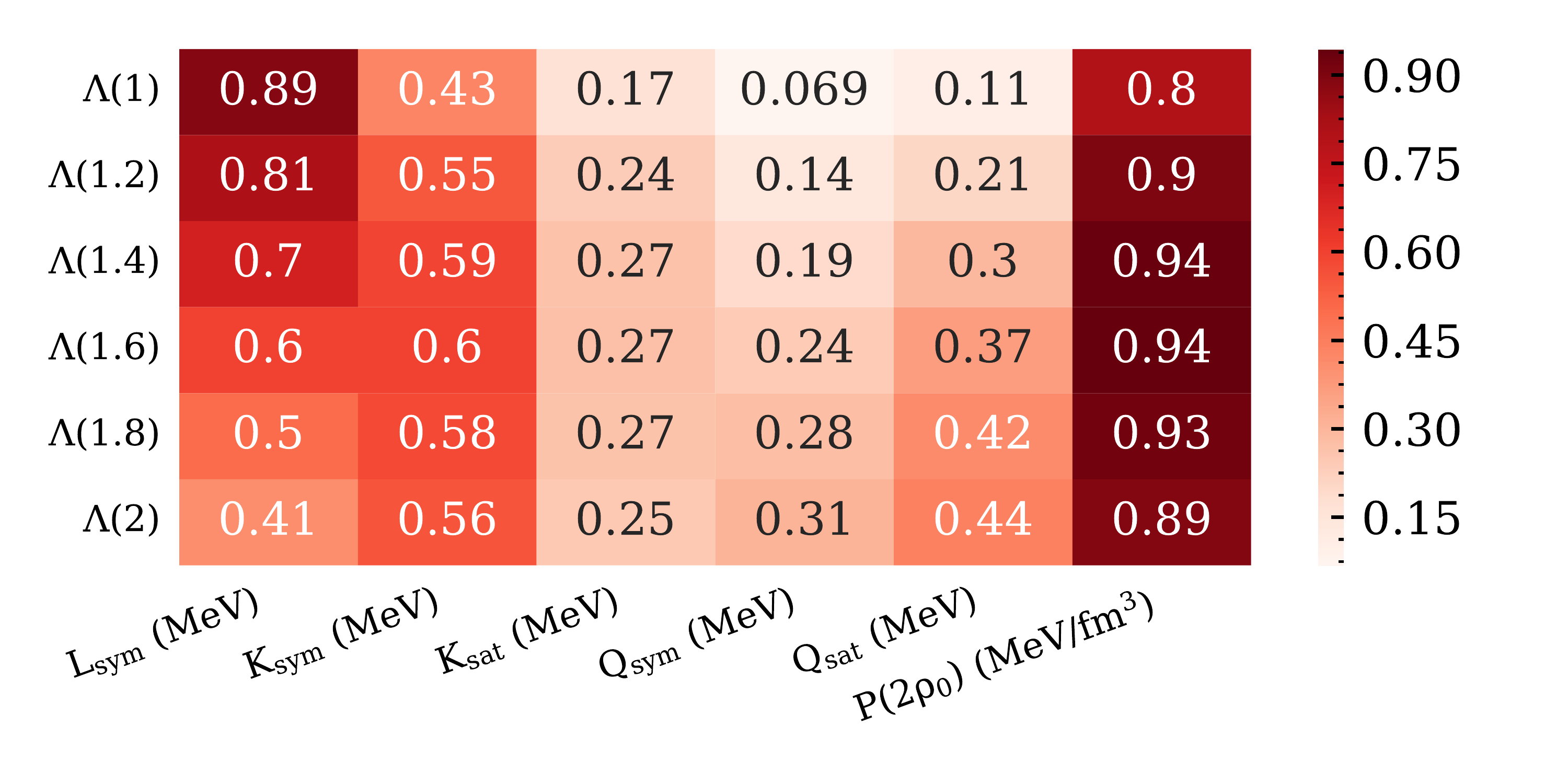}
\caption{Pearson correlation parameters for different NS masses.}
\label{CorHeatMap}
\end{figure}

To quantify this dependence of sensitivity on mass, the Pearson correlation coefficients for a few selected Taylor parameter pairs are shown in Fig.~\ref{CorHeatMap}. A gradual reduction in correlation between $L_\text{sym}$ and tidal deformability is observed as the mass of a NS increases. This is expected as relevant average density for more massive stars shift upward and away from those most directly impacted by $L_\text{sym}$. A high density parameter $P(2\rho_0)$, the pressure for pure neutron matter at twice the saturation density, is also included in Figs.~\ref{Cor} and~\ref{CorHeatMap}. The strong correlation between tidal deformability and $P(2\rho_0)$ is consistent with prior work~\cite{Abbott2018,Lim2018,Lattimer2001,Tews2018}. While this strong correlation is maintained for both heavy and light NS, the slope of the correlation becomes smaller reflecting the decrease in average values and variations of $\Lambda$ with stellar mass. 

\begin{figure}
\includegraphics[width=0.9\linewidth]{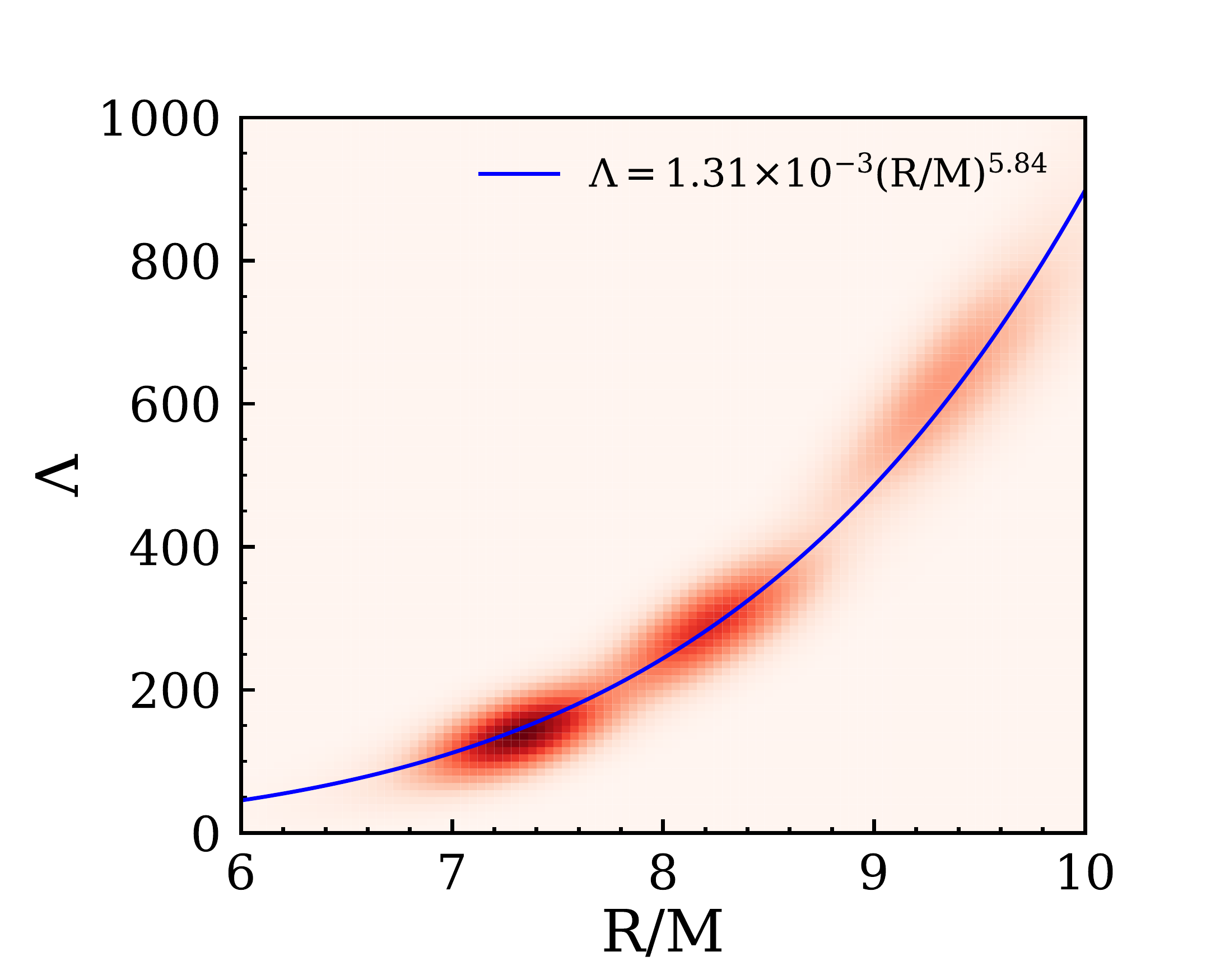}
\caption{Tidal deformability vs. inverse compactness for 1.2, 1.4, 1.6, 1.8 solar mass NS.}
\label{Compactness}
\end{figure}

Such decrease is correlated with an increase in stellar compactness. Using the posterior probability distributions for the Taylor expansion parameters, we can also make predictions on the relation between stellar mass and inverse compactness ($R/M$). Fig.~\ref{Compactness} shows tidal deformability plotted against inverse compactness, with calculation results for 1.2, 1.4, 1.6 and 1.8 solar mass NS all combined together. It is consistent with Eq.~\eqref{lambda} where $\Lambda \propto k_2(R/M )^5$. The best fitted power law has an index of 5.84 due to additional interdependence of tidal Love number $k_2$ and $R/M$. The result is consistent with Refs.~\cite{Piekarewicz2019, Maselli2013, Lattimer2019}.

After submission of this work for publication, we found that independently and in parallel, Ref.~\cite{Guven2020} conducts a very similar analysis using ELFc. Our work examines correlations between more parameters and our study extends to higher mass neutron star.  Ref.~\cite{Guven2020} uses much wider priors while our prior is more restrictive and provide finer details in a smaller phase-space. In addition, they apply additional constraints on the EOS using data  from $\chi$EFT  approach and  ISGMR  collective  mode. Even though their extracted $Q_\text{sat}$ and $K_\text{sym}$ values are consistent with our extracted values, details in the correlations are not the same. The subtle differences suggest that Bayesian analysis results depend on the choice of priors and constraints applied to the EOS. 

\section{\label{Conclusion}Conclusion}

In this paper, ELFc form of Metamodeling is used in a sensitivity study of NS tidal deformability to Taylor expansion parameters of the nuclear equation of state. Constraints on the isoscalar parameters, such as $K_\text{sat}$ are found to be less affected by NS properties. For the isovector parameters, $L_\text{sym}$ is found to be most correlated with tidal deformability, closely followed by $K_\text{sym}$, although the importance of the former dwindles and reverses as NS mass increases to above 1.6 solar mass.

We have further demonstrated the global relation between tidal deformability and compactness of NS with different masses. When more merger events involving different NS masses are observed in the future, one can verify the relation of tidal deformability and inverse compactness in Fig.~\ref{Compactness} and to provide independent constraints on $L_\text{sym}$ and $K_\text{sym}$. 

A strong correlation with the pressure of matter at $2\rho_0$ is observed. This highlights the need for high-density observables from nuclear physics, as constraints on tidal deformability can be tightened with accurate high density observations. 
%The analysis of flow experiments~\cite{Danielewicz2002} and the Kaon data~\cite{FUCHS2006, LYNCH2009} constrained pressure of symmetric matter at density from $1.2-2\rho_0$, but it relied on theoretical assumptions about the symmetry energy to extrapolate pressure to pure neutron matter. Experimental programs of the S$\pi$RIT collaboration~\cite{Shane2015}, neutron and proton yield and flow data from HiRA collaboration and GSI~\cite{Russotto2016}, are all focused on obtaining robust constraints on the symmetry energy term at high density. 
A strong experimental constraint on pressure for PNM at 2$\rho_0$ complements pressure~\cite{Shane2015, Russotto2016} constraints from future measurement of NS mergers.

\section{Acknowledgments}

This work was partly supported by the US National Science Foundation under Grant PHY-1565546 and by the U.S. Department of Energy (Office of Science) under Grants DE-SC0014530, DE-NA0002923 and DE-SC001920. All the NS model calculations with High Performance Computers were performed at the Institute for Cyber Enabled Research Center at Michigan State University.

\BeforeBeginEnvironment{appendices}{\clearpage}
\appendix
\section{\label{TOVEquation}TOV equation}

The Tolman–Oppenheimer–Volkoff (TOV) equation set predicts the structure of a static spherical object under general relativity for any given EOS. The equations are:

\begin{equation}
\begin{split}
\frac{dP(r)}{dr} &= -\frac{(\mathcal{E}(r)+P(r))(M(r)+4\pi r^3P(r))}{r^2(1-2M(r)/r)}, \\
\frac{dM(r)}{dr} &= 4\pi r^2\mathcal{E}(r).
\end{split}
\end{equation}
Here geometrized units $G=c=1$ are used, $\mathcal{E}(r)$ is the energy density given by EOS, $P(r)$ is the internal pressure at given depth and $M(r)$ is the integral of gravitational mass from the core up to radius $r$. The surface is defined as the radial distance $R$ at which $P(R) = 0$. 

A list of equations whose solutions will lead to the value of $\Lambda$ from the above structural functions will be shown without derivation. Please refer to Refs.~\cite{Postnikov2010, Fattoyev2013} for details. To begin with, an auxiliary variable $y_R = y(R)$ is calculated,
\begin{equation}
r\frac{dy(r)}{dr} + y(r)^2 + y(r)F(r) + r^2Q(r) = 0.
\end{equation}
where
\begin{equation}
F(r) = \frac{r - 4\pi r^3(\mathcal{E}(r) - P(r))}{r - 2M(r)}.
\end{equation}
\begin{equation}
\begin{split}
Q(r) = &\frac{4\pi r(5\mathcal{E}(r)+9P(r)+ \frac{\mathcal{E}+P(r)}{\partial P(r)/\partial \mathcal{E}} - \frac{6}{4\pi r^2}}{r - 2M(r)} \\
&-4\Big[\frac{(M(r)+4\pi r^3P(r)}{r^2(1-2M(r)/r)}\Big]^2.
\end{split}
\end{equation}
The tidal Love number $k_2$ can then be calculated with the following expression:

\begin{equation}
\begin{split}
k_2 =& \frac{1}{20}\Big(\frac{R_s}{R}\Big)^5\big(1 - \frac{R_s}{R}\Big)^2\Big[2-y_R+(y_R-1)\frac{R_s}{R}\Big] \\
&\times\Big\{\frac{R_s}{R}\Big(6-3y_R+\frac{3R_s}{2R}(5y_R-8)+\frac{1}{4}\Big(\frac{R_s}{R}\Big)^2 \\
&\times\Big[26 - 22y_R+\frac{R_s(3y_R-2)}{R} + \Big(\frac{R_S}{R}\Big)^2(1+y_R)\Big]\Big) \\
& + 3\Big(1-\frac{R_s}{R}\Big)^2\Big[2-y_R+\frac{R_s(y_R-1)}{R}\Big] \\
& \times \ln\Big(1-\frac{R_S}{R}\Big)\Big\}^{-1}.
\end{split}
\end{equation}
In the above, $R_S=2M$ is the Schwarzschild radius. The value of $\Lambda$ is then extracted with Eq.~\eqref{lambda}.

\section{\label{mm_mapping}Metamodeling parameters and Taylor parameters mapping}

ELFc energy functional is written as a sum of kinetic energy term and potential energy term:
\begin{equation}
E_{EFLc}(\rho,\delta)=t^{FG*}(\rho, \delta) + v^N_{EFLc}(\rho, \delta),
\label{EFLc}
\end{equation}
where $\rho$ is the density and $\delta$ is the asymmetry parameter. The kinetic energy term $t^{FG*}(\rho, \delta)$ in the above is written as:
\begin{equation}
\begin{split}
t^{FG*}(\rho, \delta) = \frac{t^{FG}_{\text{sat}}}{2}\Big(\frac{\rho}{\rho_0}\Big)^{\frac{2}{3}}\Big(\Big(1+\frac{\kappa_{\text{sat}}\rho}{\rho_0}\Big)((1+\delta)^{\frac{5}{3}}+\\ (1-\delta)^{\frac{5}{3}}\Big) + \frac{\kappa_{\text{sym}}\rho}{\rho_0}\delta((1 + \delta)^{\frac{5}{3}}-(1-\delta)^{\frac{5}{3}})).
\end{split}
\label{kinetic}
\end{equation}
In the above, the parameters $t^{FG}_{\text{sat}}=\SI{22.1}{MeV}$ while $\kappa_{sym}$ and $\kappa_{sat}$ are effective mass parameters described in Eq.~\eqref{kappaform}.

The potential energy term $v^N_{EFLc}(\rho, \delta)$ is written as:
\begin{equation}
\begin{split}
v^N_{EFLc}(\rho, \delta) &= \sum^4_{i=0}\frac{1}{i!}(v^{is}_i+v^{iv}_i\delta^2)(1-(-3)^{5-i}) \\
&\times \exp\Big(-\frac{6.93\rho}{\rho_0}\Big)x^i.
\end{split}
\label{potential}
\end{equation}
In the above, the parameters $v_i^{is}$ and $v^{iv}_i$ are free parameters. These 10 parameters can be uniquely mapped onto Taylor parameters using the following formulas (For a detailed derivation, please refer to Ref.~\cite{Margueron2018}):
 
\begin{equation}
v^{is}_0 = E_{\text{sat}}-t^{FG}_{\text{sat}}(1+\kappa_{\text{sat}}),
\end{equation}
\begin{equation}
v^{is}_1 = -t^{FG}_{\text{sat}}(2+5\kappa_{\text{sat}}),
\end{equation}
\begin{equation}
v^{is}_2 = K_{\text{sat}} - 2t^{FG}_{\text{sat}}(-1+5\kappa_{\text{sat}}),
\end{equation}
\begin{equation}
v^{is}_3 = Q_{\text{sat}} - 2t^{FG}_{\text{sat}}(4-5\kappa_{\text{sat}}),
\end{equation}
\begin{equation}
v^{is}_4 = Z_{\text{sat}} - 8t^{FG}_{\text{sat}}(-7+5\kappa_{\text{sat}}),
\end{equation}
\begin{equation}
v^{iv}_0 = S_0 - \frac{5}{9}t^{FG}_{\text{sat}}(1+(\kappa_{\text{sat}} + 3\kappa_{\text{sym}})),
\end{equation}
\begin{equation}
v^{iv}_1 = L - \frac{5}{9}t^{FG}_{\text{sat}}(2+5(\kappa_{\text{sat}}+3\kappa_{\text{sym}})),
\end{equation}
\begin{equation}
v^{iv}_2 = K_{\text{sym}} - \frac{10}{9}t^{FG}_{\text{sat}}(-1+5(\kappa_{\text{sat}}+3\kappa_{\text{sym}})),
\end{equation}
\begin{equation}
v^{iv}_3 = Q_{\text{sym}} - \frac{10}{9}t^{FG}_{\text{sat}}(4-5(\kappa_{\text{sat}} + 3\kappa_{\text{sym}})),
\end{equation}
\begin{equation}
v^{iv}_4 = Z_{\text{sym}} - \frac{40}{9}t^{FG}_{\text{sat}}(-7+5(\kappa_{\text{sat}}+3\kappa_{\text{sym}})).
\end{equation}
When exploring the parameter space, Taylor parameters will be translated to Metamodeling EOS using the above formulas and NS features will then be calculated with TOV equation. Neutron star properties will be examined in order to search for Taylor parameter spaces flavored by the observed tidal deformability.

\section{\label{secfullcor}Full correlation between tidal deformability and parameters}

\begin{figure*}
\includegraphics[width=0.95\linewidth]{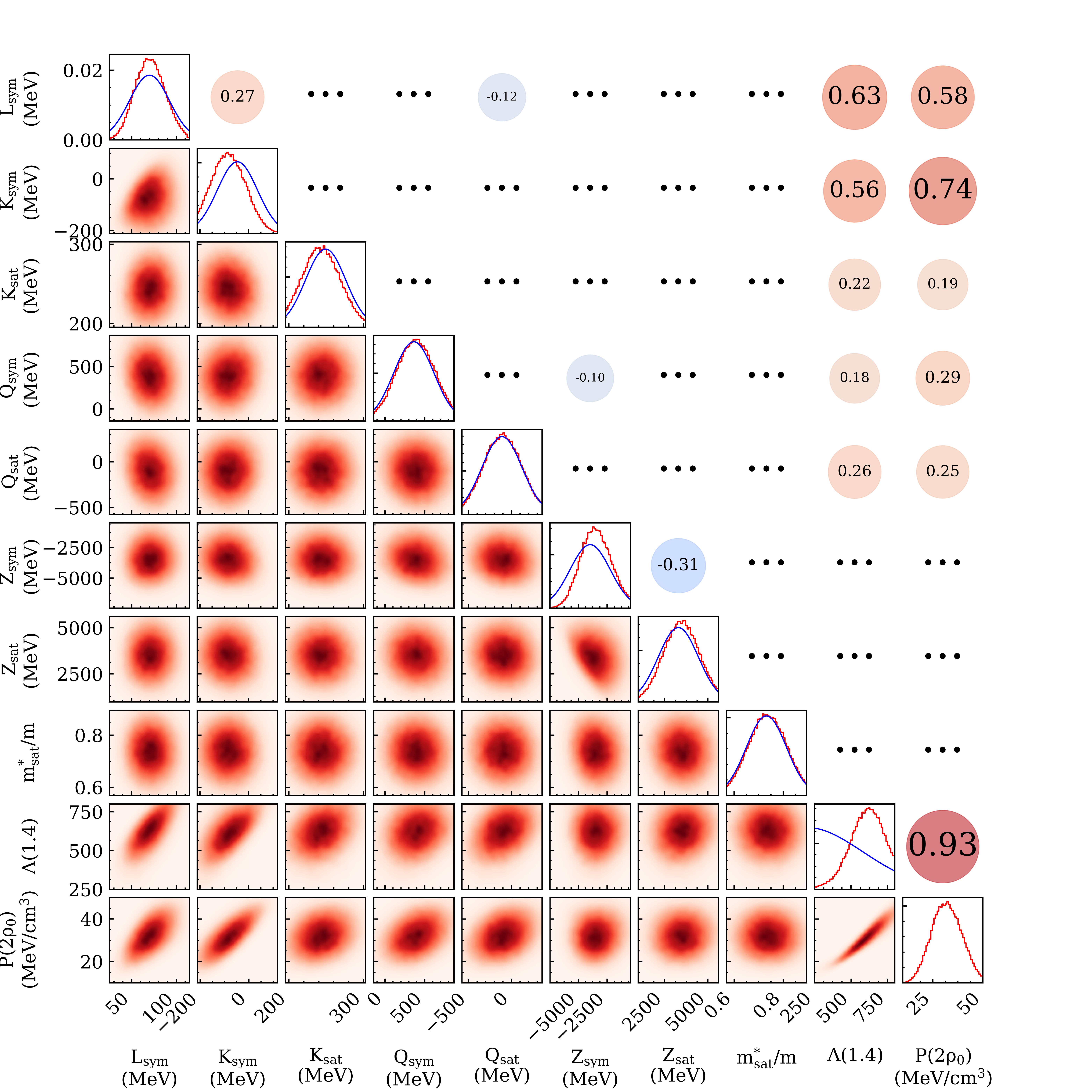}
\caption{Bivariate characteristics of posterior likelihood distributions. This is an extension to Fig.~\ref{Correlation} and correlation pairs of all parameters pair are shown. Three regions can be distinguished. The lower triangle panels show likelihood distributions, with intensity proportional to distribution value, for pairs of Taylor parameters. The diagonal panels display marginalized distribution for each parameter. The upper triangular region shows Pearson correlation coefficient for parameter pairs, but when correlation in magnitude is less than 0.1, it is omitted and 3 dots are put in place of its value.}
\label{FullCor}
\end{figure*}

The correlation between $L_{\text{sym}}$, $K_{\text{sym}}$, $K_{\text{sat}}$, $Q_{\text{sym}}$, $Q_{\text{sat}}$,$Z_{\text{sym}}$, $Z_{\text{sat}}$, $\Big(m_{\text{sat}}/m\Big)$, $P(2\rho_0)$ and $\Lambda$ are shown in Fig.~\ref{FullCor}. This is an extension of Fig.~\ref{Cor} where bivariate distributions of some selected parameters are shown. The organization is similar: Lower triangles show bivariate distributions between variables and marginal distribution of each variable is shown on the diagonal. The upper triangles shows Pearson correlation coefficients between each variable pairs if it is larger than 0.1 otherwise they are omitted for simplicity and 3 dots are put in its place.


\begin{thebibliography}{97}%
\makeatletter
\providecommand \@ifxundefined [1]{%
 \@ifx{#1\undefined}
}%
\providecommand \@ifnum [1]{%
 \ifnum #1\expandafter \@firstoftwo
 \else \expandafter \@secondoftwo
 \fi
}%
\providecommand \@ifx [1]{%
 \ifx #1\expandafter \@firstoftwo
 \else \expandafter \@secondoftwo
 \fi
}%
\providecommand \natexlab [1]{#1}%
\providecommand \enquote  [1]{``#1''}%
\providecommand \bibnamefont  [1]{#1}%
\providecommand \bibfnamefont [1]{#1}%
\providecommand \citenamefont [1]{#1}%
\providecommand \href@noop [0]{\@secondoftwo}%
\providecommand \href [0]{\begingroup \@sanitize@url \@href}%
\providecommand \@href[1]{\@@startlink{#1}\@@href}%
\providecommand \@@href[1]{\endgroup#1\@@endlink}%
\providecommand \@sanitize@url [0]{\catcode `\\12\catcode `\$12\catcode
  `\&12\catcode `\#12\catcode `\^12\catcode `\_12\catcode `\%12\relax}%
\providecommand \@@startlink[1]{}%
\providecommand \@@endlink[0]{}%
\providecommand \url  [0]{\begingroup\@sanitize@url \@url }%
\providecommand \@url [1]{\endgroup\@href {#1}{\urlprefix }}%
\providecommand \urlprefix  [0]{URL }%
\providecommand \Eprint [0]{\href }%
\providecommand \doibase [0]{http://dx.doi.org/}%
\providecommand \selectlanguage [0]{\@gobble}%
\providecommand \bibinfo  [0]{\@secondoftwo}%
\providecommand \bibfield  [0]{\@secondoftwo}%
\providecommand \translation [1]{[#1]}%
\providecommand \BibitemOpen [0]{}%
\providecommand \bibitemStop [0]{}%
\providecommand \bibitemNoStop [0]{.\EOS\space}%
\providecommand \EOS [0]{\spacefactor3000\relax}%
\providecommand \BibitemShut  [1]{\csname bibitem#1\endcsname}%
\let\auto@bib@innerbib\@empty
%</preamble>
\bibitem [{\citenamefont {Li}\ \emph {et~al.}(2019{\natexlab{a}})\citenamefont
  {Li}, \citenamefont {Krastev}, \citenamefont {Wen},\ and\ \citenamefont
  {Zhang}}]{Li2019}%
  \BibitemOpen
  \bibfield  {author} {\bibinfo {author} {\bibfnamefont {B.-A.}\ \bibnamefont
  {Li}}, \bibinfo {author} {\bibfnamefont {P.~G.}\ \bibnamefont {Krastev}},
  \bibinfo {author} {\bibfnamefont {D.-H.}\ \bibnamefont {Wen}}, \ and\
  \bibinfo {author} {\bibfnamefont {N.-B.}\ \bibnamefont {Zhang}},\ }\href
  {\doibase 10.1140/epja/i2019-12780-8} {\bibfield  {journal} {\bibinfo
  {journal} {The European Physical Journal A}\ }\textbf {\bibinfo {volume}
  {55}},\ \bibinfo {pages} {117} (\bibinfo {year}
  {2019}{\natexlab{a}})}\BibitemShut {NoStop}%
\bibitem [{\citenamefont {Baldo}\ and\ \citenamefont
  {Burgio}(2016)}]{Baldo2016}%
  \BibitemOpen
  \bibfield  {author} {\bibinfo {author} {\bibfnamefont {M.}~\bibnamefont
  {Baldo}}\ and\ \bibinfo {author} {\bibfnamefont {G.}~\bibnamefont {Burgio}},\
  }\href {\doibase 10.1016/j.ppnp.2016.06.006} {\bibfield  {journal} {\bibinfo
  {journal} {Progress in Particle and Nuclear Physics}\ }\textbf {\bibinfo
  {volume} {91}} (\bibinfo {year} {2016}),\
  10.1016/j.ppnp.2016.06.006}\BibitemShut {NoStop}%
\bibitem [{\citenamefont {Lattimer}(2012)}]{Lattimer2012}%
  \BibitemOpen
  \bibfield  {author} {\bibinfo {author} {\bibfnamefont {J.~M.}\ \bibnamefont
  {Lattimer}},\ }\href {\doibase 10.1146/annurev-nucl-102711-095018} {\bibfield
   {journal} {\bibinfo  {journal} {Annual Review of Nuclear and Particle
  Science}\ }\textbf {\bibinfo {volume} {62}},\ \bibinfo {pages} {485}
  (\bibinfo {year} {2012})},\ \Eprint
  {http://arxiv.org/abs/https://doi.org/10.1146/annurev-nucl-102711-095018}
  {https://doi.org/10.1146/annurev-nucl-102711-095018} \BibitemShut {NoStop}%
\bibitem [{\citenamefont {Holt}\ and\ \citenamefont {Kaiser}(2017)}]{Holt2017}%
  \BibitemOpen
  \bibfield  {author} {\bibinfo {author} {\bibfnamefont {J.~W.}\ \bibnamefont
  {Holt}}\ and\ \bibinfo {author} {\bibfnamefont {N.}~\bibnamefont {Kaiser}},\
  }\href {\doibase 10.1103/PhysRevC.95.034326} {\bibfield  {journal} {\bibinfo
  {journal} {Phys. Rev. C}\ }\textbf {\bibinfo {volume} {95}},\ \bibinfo
  {pages} {034326} (\bibinfo {year} {2017})}\BibitemShut {NoStop}%
\bibitem [{\citenamefont {Tsang}\ \emph
  {et~al.}(2019{\natexlab{a}})\citenamefont {Tsang}, \citenamefont {Lynch},
  \citenamefont {Danielewicz},\ and\ \citenamefont {Tsang}}]{Tsang2019533}%
  \BibitemOpen
  \bibfield  {author} {\bibinfo {author} {\bibfnamefont {M.}~\bibnamefont
  {Tsang}}, \bibinfo {author} {\bibfnamefont {W.}~\bibnamefont {Lynch}},
  \bibinfo {author} {\bibfnamefont {P.}~\bibnamefont {Danielewicz}}, \ and\
  \bibinfo {author} {\bibfnamefont {C.}~\bibnamefont {Tsang}},\ }\href
  {\doibase https://doi.org/10.1016/j.physletb.2019.06.059} {\bibfield
  {journal} {\bibinfo  {journal} {Physics Letters B}\ }\textbf {\bibinfo
  {volume} {795}},\ \bibinfo {pages} {533 } (\bibinfo {year}
  {2019}{\natexlab{a}})}\BibitemShut {NoStop}%
\bibitem [{\citenamefont {Danielewicz}\ \emph {et~al.}(2002)\citenamefont
  {Danielewicz}, \citenamefont {Lacey},\ and\ \citenamefont
  {Lynch}}]{Danielewicz2002}%
  \BibitemOpen
  \bibfield  {author} {\bibinfo {author} {\bibfnamefont {P.}~\bibnamefont
  {Danielewicz}}, \bibinfo {author} {\bibfnamefont {R.}~\bibnamefont {Lacey}},
  \ and\ \bibinfo {author} {\bibfnamefont {W.~G.}\ \bibnamefont {Lynch}},\
  }\href {\doibase 10.1126/science.1078070} {\bibfield  {journal} {\bibinfo
  {journal} {Science}\ }\textbf {\bibinfo {volume} {298}},\ \bibinfo {pages}
  {1592} (\bibinfo {year} {2002})}\BibitemShut {NoStop}%
\bibitem [{\citenamefont {Fuchs}(2006)}]{FUCHS2006}%
  \BibitemOpen
  \bibfield  {author} {\bibinfo {author} {\bibfnamefont {C.}~\bibnamefont
  {Fuchs}},\ }\href {\doibase https://doi.org/10.1016/j.ppnp.2005.07.004}
  {\bibfield  {journal} {\bibinfo  {journal} {Progress in Particle and Nuclear
  Physics}\ }\textbf {\bibinfo {volume} {56}},\ \bibinfo {pages} {1 } (\bibinfo
  {year} {2006})}\BibitemShut {NoStop}%
\bibitem [{\citenamefont {Lynch}\ \emph {et~al.}(2009)\citenamefont {Lynch},
  \citenamefont {Tsang}, \citenamefont {Zhang}, \citenamefont {Danielewicz},
  \citenamefont {Famiano}, \citenamefont {Li},\ and\ \citenamefont
  {Steiner}}]{LYNCH2009}%
  \BibitemOpen
  \bibfield  {author} {\bibinfo {author} {\bibfnamefont {W.~G.}\ \bibnamefont
  {Lynch}}, \bibinfo {author} {\bibfnamefont {M.~B.}\ \bibnamefont {Tsang}},
  \bibinfo {author} {\bibfnamefont {Y.}~\bibnamefont {Zhang}}, \bibinfo
  {author} {\bibfnamefont {P.}~\bibnamefont {Danielewicz}}, \bibinfo {author}
  {\bibfnamefont {M.}~\bibnamefont {Famiano}}, \bibinfo {author} {\bibfnamefont
  {Z.}~\bibnamefont {Li}}, \ and\ \bibinfo {author} {\bibfnamefont {A.~W.}\
  \bibnamefont {Steiner}},\ }\href {\doibase
  https://doi.org/10.1016/j.ppnp.2009.01.001} {\bibfield  {journal} {\bibinfo
  {journal} {Progress in Particle and Nuclear Physics}\ }\textbf {\bibinfo
  {volume} {62}},\ \bibinfo {pages} {427 } (\bibinfo {year} {2009})},\ \bibinfo
  {note} {heavy-Ion Collisions from the Coulomb Barrier to the Quark-Gluon
  Plasma}\BibitemShut {NoStop}%
\bibitem [{\citenamefont {F{\`e}vre}\ \emph {et~al.}(2016)\citenamefont
  {F{\`e}vre}, \citenamefont {Leifels}, \citenamefont {Reisdorf}, \citenamefont
  {Aichelin},\ and\ \citenamefont {Hartnack}}]{LEFEVRE2016}%
  \BibitemOpen
  \bibfield  {author} {\bibinfo {author} {\bibfnamefont {A.~L.}\ \bibnamefont
  {F{\`e}vre}}, \bibinfo {author} {\bibfnamefont {Y.}~\bibnamefont {Leifels}},
  \bibinfo {author} {\bibfnamefont {W.}~\bibnamefont {Reisdorf}}, \bibinfo
  {author} {\bibfnamefont {J.}~\bibnamefont {Aichelin}}, \ and\ \bibinfo
  {author} {\bibfnamefont {C.}~\bibnamefont {Hartnack}},\ }\href {\doibase
  https://doi.org/10.1016/j.nuclphysa.2015.09.015} {\bibfield  {journal}
  {\bibinfo  {journal} {Nuclear Physics A}\ }\textbf {\bibinfo {volume}
  {945}},\ \bibinfo {pages} {112 } (\bibinfo {year} {2016})}\BibitemShut
  {NoStop}%
\bibitem [{\citenamefont {Steiner}\ \emph {et~al.}(2013)\citenamefont
  {Steiner}, \citenamefont {Lattimer},\ and\ \citenamefont
  {Brown}}]{Steiner2013}%
  \BibitemOpen
  \bibfield  {author} {\bibinfo {author} {\bibfnamefont {A.~W.}\ \bibnamefont
  {Steiner}}, \bibinfo {author} {\bibfnamefont {J.~M.}\ \bibnamefont
  {Lattimer}}, \ and\ \bibinfo {author} {\bibfnamefont {E.~F.}\ \bibnamefont
  {Brown}},\ }\href {\doibase 10.1088/2041-8205/765/1/l5} {\bibfield  {journal}
  {\bibinfo  {journal} {The Astrophysical Journal}\ }\textbf {\bibinfo {volume}
  {765}},\ \bibinfo {pages} {L5} (\bibinfo {year} {2013})}\BibitemShut
  {NoStop}%
\bibitem [{\citenamefont {Abbott}\ \emph {et~al.}(2017)\citenamefont {Abbott}
  \emph {et~al.}}]{Abbott2017}%
  \BibitemOpen
  \bibfield  {author} {\bibinfo {author} {\bibfnamefont {B.~P.}\ \bibnamefont
  {Abbott}} \emph {et~al.} (\bibinfo {collaboration} {LIGO Scientific
  Collaboration and Virgo Collaboration}),\ }\href {\doibase
  10.1103/PhysRevLett.119.161101} {\bibfield  {journal} {\bibinfo  {journal}
  {Phys. Rev. Lett.}\ }\textbf {\bibinfo {volume} {119}},\ \bibinfo {pages}
  {161101} (\bibinfo {year} {2017})}\BibitemShut {NoStop}%
\bibitem [{\citenamefont {Damour}\ \emph {et~al.}(1992)\citenamefont {Damour},
  \citenamefont {Soffel},\ and\ \citenamefont {Xu}}]{Damour1992}%
  \BibitemOpen
  \bibfield  {author} {\bibinfo {author} {\bibfnamefont {T.}~\bibnamefont
  {Damour}}, \bibinfo {author} {\bibfnamefont {M.}~\bibnamefont {Soffel}}, \
  and\ \bibinfo {author} {\bibfnamefont {C.}~\bibnamefont {Xu}},\ }\href
  {\doibase 10.1103/PhysRevD.45.1017} {\bibfield  {journal} {\bibinfo
  {journal} {Phys. Rev. D}\ }\textbf {\bibinfo {volume} {45}},\ \bibinfo
  {pages} {1017} (\bibinfo {year} {1992})}\BibitemShut {NoStop}%
\bibitem [{\citenamefont {Flanagan}\ and\ \citenamefont
  {Hinderer}(2008)}]{Flanagan2008}%
  \BibitemOpen
  \bibfield  {author} {\bibinfo {author} {\bibfnamefont {E.~E.}\ \bibnamefont
  {Flanagan}}\ and\ \bibinfo {author} {\bibfnamefont {T.}~\bibnamefont
  {Hinderer}},\ }\href {\doibase 10.1103/PhysRevD.77.021502} {\bibfield
  {journal} {\bibinfo  {journal} {Phys. Rev. D}\ }\textbf {\bibinfo {volume}
  {77}},\ \bibinfo {pages} {021502} (\bibinfo {year} {2008})}\BibitemShut
  {NoStop}%
\bibitem [{\citenamefont {Damour}\ \emph {et~al.}(2012)\citenamefont {Damour},
  \citenamefont {Nagar},\ and\ \citenamefont {Villain}}]{Damour2012}%
  \BibitemOpen
  \bibfield  {author} {\bibinfo {author} {\bibfnamefont {T.}~\bibnamefont
  {Damour}}, \bibinfo {author} {\bibfnamefont {A.}~\bibnamefont {Nagar}}, \
  and\ \bibinfo {author} {\bibfnamefont {L.}~\bibnamefont {Villain}},\ }\href
  {\doibase 10.1103/PhysRevD.85.123007} {\bibfield  {journal} {\bibinfo
  {journal} {Phys. Rev. D}\ }\textbf {\bibinfo {volume} {85}},\ \bibinfo
  {pages} {123007} (\bibinfo {year} {2012})}\BibitemShut {NoStop}%
\bibitem [{\citenamefont {Binnington}\ and\ \citenamefont
  {Poisson}(2009)}]{Binnington2009}%
  \BibitemOpen
  \bibfield  {author} {\bibinfo {author} {\bibfnamefont {T.}~\bibnamefont
  {Binnington}}\ and\ \bibinfo {author} {\bibfnamefont {E.}~\bibnamefont
  {Poisson}},\ }\href {\doibase 10.1103/PhysRevD.80.084018} {\bibfield
  {journal} {\bibinfo  {journal} {Phys. Rev. D}\ }\textbf {\bibinfo {volume}
  {80}},\ \bibinfo {pages} {084018} (\bibinfo {year} {2009})}\BibitemShut
  {NoStop}%
\bibitem [{\citenamefont {Postnikov}\ \emph {et~al.}(2010)\citenamefont
  {Postnikov}, \citenamefont {Prakash},\ and\ \citenamefont
  {Lattimer}}]{Postnikov2010}%
  \BibitemOpen
  \bibfield  {author} {\bibinfo {author} {\bibfnamefont {S.}~\bibnamefont
  {Postnikov}}, \bibinfo {author} {\bibfnamefont {M.}~\bibnamefont {Prakash}},
  \ and\ \bibinfo {author} {\bibfnamefont {J.~M.}\ \bibnamefont {Lattimer}},\
  }\href {\doibase 10.1103/PhysRevD.82.024016} {\bibfield  {journal} {\bibinfo
  {journal} {Phys. Rev. D}\ }\textbf {\bibinfo {volume} {82}},\ \bibinfo
  {pages} {024016} (\bibinfo {year} {2010})}\BibitemShut {NoStop}%
\bibitem [{\citenamefont {Piekarewicz}\ and\ \citenamefont
  {Fattoyev}(2019)}]{Piekarewicz2019}%
  \BibitemOpen
  \bibfield  {author} {\bibinfo {author} {\bibfnamefont {J.}~\bibnamefont
  {Piekarewicz}}\ and\ \bibinfo {author} {\bibfnamefont {F.~J.}\ \bibnamefont
  {Fattoyev}},\ }\href {\doibase 10.1103/PhysRevC.99.045802} {\bibfield
  {journal} {\bibinfo  {journal} {Phys. Rev. C}\ }\textbf {\bibinfo {volume}
  {99}},\ \bibinfo {pages} {045802} (\bibinfo {year} {2019})}\BibitemShut
  {NoStop}%
\bibitem [{\citenamefont {Abbott}\ \emph {et~al.}(2018)\citenamefont {Abbott}
  \emph {et~al.}}]{Abbott2018}%
  \BibitemOpen
  \bibfield  {author} {\bibinfo {author} {\bibfnamefont {B.~P.}\ \bibnamefont
  {Abbott}} \emph {et~al.} (\bibinfo {collaboration} {The LIGO Scientific
  Collaboration and the Virgo Collaboration}),\ }\href {\doibase
  10.1103/PhysRevLett.121.161101} {\bibfield  {journal} {\bibinfo  {journal}
  {Phys. Rev. Lett.}\ }\textbf {\bibinfo {volume} {121}},\ \bibinfo {pages}
  {161101} (\bibinfo {year} {2018})}\BibitemShut {NoStop}%
\bibitem [{\citenamefont {Kortelainen}\ \emph {et~al.}(2010)\citenamefont
  {Kortelainen}, \citenamefont {Lesinski}, \citenamefont {Mor\'e},
  \citenamefont {Nazarewicz}, \citenamefont {Sarich}, \citenamefont {Schunck},
  \citenamefont {Stoitsov},\ and\ \citenamefont {Wild}}]{Kortelainen2010}%
  \BibitemOpen
  \bibfield  {author} {\bibinfo {author} {\bibfnamefont {M.}~\bibnamefont
  {Kortelainen}}, \bibinfo {author} {\bibfnamefont {T.}~\bibnamefont
  {Lesinski}}, \bibinfo {author} {\bibfnamefont {J.}~\bibnamefont {Mor\'e}},
  \bibinfo {author} {\bibfnamefont {W.}~\bibnamefont {Nazarewicz}}, \bibinfo
  {author} {\bibfnamefont {J.}~\bibnamefont {Sarich}}, \bibinfo {author}
  {\bibfnamefont {N.}~\bibnamefont {Schunck}}, \bibinfo {author} {\bibfnamefont
  {M.~V.}\ \bibnamefont {Stoitsov}}, \ and\ \bibinfo {author} {\bibfnamefont
  {S.}~\bibnamefont {Wild}},\ }\href {\doibase 10.1103/PhysRevC.82.024313}
  {\bibfield  {journal} {\bibinfo  {journal} {Phys. Rev. C}\ }\textbf {\bibinfo
  {volume} {82}},\ \bibinfo {pages} {024313} (\bibinfo {year}
  {2010})}\BibitemShut {NoStop}%
\bibitem [{\citenamefont {Brown}(2013)}]{Brown2013}%
  \BibitemOpen
  \bibfield  {author} {\bibinfo {author} {\bibfnamefont {B.~A.}\ \bibnamefont
  {Brown}},\ }\href {\doibase 10.1103/PhysRevLett.111.232502} {\bibfield
  {journal} {\bibinfo  {journal} {Phys. Rev. Lett.}\ }\textbf {\bibinfo
  {volume} {111}},\ \bibinfo {pages} {232502} (\bibinfo {year}
  {2013})}\BibitemShut {NoStop}%
\bibitem [{\citenamefont {Zhang}\ and\ \citenamefont {Chen}(2013)}]{Zhang2013}%
  \BibitemOpen
  \bibfield  {author} {\bibinfo {author} {\bibfnamefont {Z.}~\bibnamefont
  {Zhang}}\ and\ \bibinfo {author} {\bibfnamefont {L.-W.}\ \bibnamefont
  {Chen}},\ }\href {\doibase 10.1016/j.physletb.2013.08.002} {\bibfield
  {journal} {\bibinfo  {journal} {Phys. Lett.}\ }\textbf {\bibinfo {volume}
  {B726}},\ \bibinfo {pages} {234} (\bibinfo {year} {2013})},\ \Eprint
  {http://arxiv.org/abs/1302.5327} {arXiv:1302.5327 [nucl-th]} \BibitemShut
  {NoStop}%
%%CITATION = ARXIV:1302.5327;%%
\bibitem [{\citenamefont {Danielewicz}\ \emph {et~al.}(2017)\citenamefont
  {Danielewicz}, \citenamefont {Singh},\ and\ \citenamefont
  {Lee}}]{Danielewicz2016}%
  \BibitemOpen
  \bibfield  {author} {\bibinfo {author} {\bibfnamefont {P.}~\bibnamefont
  {Danielewicz}}, \bibinfo {author} {\bibfnamefont {P.}~\bibnamefont {Singh}},
  \ and\ \bibinfo {author} {\bibfnamefont {J.}~\bibnamefont {Lee}},\ }\href
  {\doibase 10.1016/j.nuclphysa.2016.11.008} {\bibfield  {journal} {\bibinfo
  {journal} {Nucl. Phys.}\ }\textbf {\bibinfo {volume} {A958}},\ \bibinfo
  {pages} {147} (\bibinfo {year} {2017})},\ \Eprint
  {http://arxiv.org/abs/1611.01871} {arXiv:1611.01871 [nucl-th]} \BibitemShut
  {NoStop}%
%%CITATION = ARXIV:1611.01871;%%
\bibitem [{\citenamefont {Tsang}\ \emph
  {et~al.}(2019{\natexlab{b}})\citenamefont {Tsang}, \citenamefont {Tsang},
  \citenamefont {Danielewicz}, \citenamefont {Fattoyev},\ and\ \citenamefont
  {Lynch}}]{TSANG2019}%
  \BibitemOpen
  \bibfield  {author} {\bibinfo {author} {\bibfnamefont {C.~Y.}\ \bibnamefont
  {Tsang}}, \bibinfo {author} {\bibfnamefont {M.~B.}\ \bibnamefont {Tsang}},
  \bibinfo {author} {\bibfnamefont {P.}~\bibnamefont {Danielewicz}}, \bibinfo
  {author} {\bibfnamefont {F.~J.}\ \bibnamefont {Fattoyev}}, \ and\ \bibinfo
  {author} {\bibfnamefont {W.~G.}\ \bibnamefont {Lynch}},\ }\href {\doibase
  https://doi.org/10.1016/j.physletb.2019.05.055} {\bibfield  {journal}
  {\bibinfo  {journal} {Physics Letters B}\ }\textbf {\bibinfo {volume}
  {796}},\ \bibinfo {pages} {1 } (\bibinfo {year}
  {2019}{\natexlab{b}})}\BibitemShut {NoStop}%
\bibitem [{\citenamefont {Malik}\ \emph {et~al.}(2018)\citenamefont {Malik},
  \citenamefont {Alam}, \citenamefont {Fortin}, \citenamefont {Provid\^encia},
  \citenamefont {Agrawal}, \citenamefont {Jha}, \citenamefont {Kumar},\ and\
  \citenamefont {Patra}}]{Malik2018}%
  \BibitemOpen
  \bibfield  {author} {\bibinfo {author} {\bibfnamefont {T.}~\bibnamefont
  {Malik}}, \bibinfo {author} {\bibfnamefont {N.}~\bibnamefont {Alam}},
  \bibinfo {author} {\bibfnamefont {M.}~\bibnamefont {Fortin}}, \bibinfo
  {author} {\bibfnamefont {C.}~\bibnamefont {Provid\^encia}}, \bibinfo {author}
  {\bibfnamefont {B.~K.}\ \bibnamefont {Agrawal}}, \bibinfo {author}
  {\bibfnamefont {T.~K.}\ \bibnamefont {Jha}}, \bibinfo {author} {\bibfnamefont
  {B.}~\bibnamefont {Kumar}}, \ and\ \bibinfo {author} {\bibfnamefont {S.~K.}\
  \bibnamefont {Patra}},\ }\href {\doibase 10.1103/PhysRevC.98.035804}
  {\bibfield  {journal} {\bibinfo  {journal} {Phys. Rev. C}\ }\textbf {\bibinfo
  {volume} {98}},\ \bibinfo {pages} {035804} (\bibinfo {year}
  {2018})}\BibitemShut {NoStop}%
\bibitem [{\citenamefont {Gil}\ \emph {et~al.}(2019)\citenamefont {Gil},
  \citenamefont {Kim}, \citenamefont {Hyun}, \citenamefont {Papakonstantinou},\
  and\ \citenamefont {Oh}}]{Gil2019}%
  \BibitemOpen
  \bibfield  {author} {\bibinfo {author} {\bibfnamefont {H.}~\bibnamefont
  {Gil}}, \bibinfo {author} {\bibfnamefont {Y.-M.}\ \bibnamefont {Kim}},
  \bibinfo {author} {\bibfnamefont {C.~H.}\ \bibnamefont {Hyun}}, \bibinfo
  {author} {\bibfnamefont {P.}~\bibnamefont {Papakonstantinou}}, \ and\
  \bibinfo {author} {\bibfnamefont {Y.}~\bibnamefont {Oh}},\ }\href {\doibase
  10.1103/PhysRevC.100.014312} {\bibfield  {journal} {\bibinfo  {journal}
  {Phys. Rev. C}\ }\textbf {\bibinfo {volume} {100}},\ \bibinfo {pages}
  {014312} (\bibinfo {year} {2019})}\BibitemShut {NoStop}%
\bibitem [{\citenamefont {Carson}\ \emph {et~al.}(2019)\citenamefont {Carson},
  \citenamefont {Steiner},\ and\ \citenamefont {Yagi}}]{Carson2019}%
  \BibitemOpen
  \bibfield  {author} {\bibinfo {author} {\bibfnamefont {Z.}~\bibnamefont
  {Carson}}, \bibinfo {author} {\bibfnamefont {A.~W.}\ \bibnamefont {Steiner}},
  \ and\ \bibinfo {author} {\bibfnamefont {K.}~\bibnamefont {Yagi}},\ }\href
  {\doibase 10.1103/PhysRevD.99.043010} {\bibfield  {journal} {\bibinfo
  {journal} {Phys. Rev. D}\ }\textbf {\bibinfo {volume} {99}},\ \bibinfo
  {pages} {043010} (\bibinfo {year} {2019})}\BibitemShut {NoStop}%
\bibitem [{\citenamefont {Lim}\ and\ \citenamefont {Holt}(2018)}]{Lim2018}%
  \BibitemOpen
  \bibfield  {author} {\bibinfo {author} {\bibfnamefont {Y.}~\bibnamefont
  {Lim}}\ and\ \bibinfo {author} {\bibfnamefont {J.~W.}\ \bibnamefont {Holt}},\
  }\href {\doibase 10.1103/PhysRevLett.121.062701} {\bibfield  {journal}
  {\bibinfo  {journal} {Phys. Rev. Lett.}\ }\textbf {\bibinfo {volume} {121}},\
  \bibinfo {pages} {062701} (\bibinfo {year} {2018})}\BibitemShut {NoStop}%
\bibitem [{\citenamefont {Lattimer}\ and\ \citenamefont
  {Prakash}(2001)}]{Lattimer2001}%
  \BibitemOpen
  \bibfield  {author} {\bibinfo {author} {\bibfnamefont {J.~M.}\ \bibnamefont
  {Lattimer}}\ and\ \bibinfo {author} {\bibfnamefont {M.}~\bibnamefont
  {Prakash}},\ }\href {\doibase 10.1086/319702} {\bibfield  {journal} {\bibinfo
   {journal} {The Astrophysical Journal}\ }\textbf {\bibinfo {volume} {550}},\
  \bibinfo {pages} {426} (\bibinfo {year} {2001})}\BibitemShut {NoStop}%
\bibitem [{\citenamefont {Dutra}\ \emph {et~al.}(2012)\citenamefont {Dutra},
  \citenamefont {Louren\ifmmode~\mbox{\c{c}}\else \c{c}\fi{}o}, \citenamefont
  {S\'a~Martins}, \citenamefont {Delfino}, \citenamefont {Stone},\ and\
  \citenamefont {Stevenson}}]{Dutra2012}%
  \BibitemOpen
  \bibfield  {author} {\bibinfo {author} {\bibfnamefont {M.}~\bibnamefont
  {Dutra}}, \bibinfo {author} {\bibfnamefont {O.}~\bibnamefont
  {Louren\ifmmode~\mbox{\c{c}}\else \c{c}\fi{}o}}, \bibinfo {author}
  {\bibfnamefont {J.~S.}\ \bibnamefont {S\'a~Martins}}, \bibinfo {author}
  {\bibfnamefont {A.}~\bibnamefont {Delfino}}, \bibinfo {author} {\bibfnamefont
  {J.~R.}\ \bibnamefont {Stone}}, \ and\ \bibinfo {author} {\bibfnamefont
  {P.~D.}\ \bibnamefont {Stevenson}},\ }\href {\doibase
  10.1103/PhysRevC.85.035201} {\bibfield  {journal} {\bibinfo  {journal} {Phys.
  Rev. C}\ }\textbf {\bibinfo {volume} {85}},\ \bibinfo {pages} {035201}
  (\bibinfo {year} {2012})}\BibitemShut {NoStop}%
\bibitem [{\citenamefont {Khan}\ \emph {et~al.}(2012)\citenamefont {Khan},
  \citenamefont {Margueron},\ and\ \citenamefont {Vida\~na}}]{Khan2012}%
  \BibitemOpen
  \bibfield  {author} {\bibinfo {author} {\bibfnamefont {E.}~\bibnamefont
  {Khan}}, \bibinfo {author} {\bibfnamefont {J.}~\bibnamefont {Margueron}}, \
  and\ \bibinfo {author} {\bibfnamefont {I.}~\bibnamefont {Vida\~na}},\ }\href
  {\doibase 10.1103/PhysRevLett.109.092501} {\bibfield  {journal} {\bibinfo
  {journal} {Phys. Rev. Lett.}\ }\textbf {\bibinfo {volume} {109}},\ \bibinfo
  {pages} {092501} (\bibinfo {year} {2012})}\BibitemShut {NoStop}%
\bibitem [{\citenamefont {Khan}(2013)}]{Khan2013}%
  \BibitemOpen
  \bibfield  {author} {\bibinfo {author} {\bibfnamefont {E.}~\bibnamefont
  {Khan}},\ }\href {\doibase 10.1088/0031-8949/2013/t152/014008} {\bibfield
  {journal} {\bibinfo  {journal} {Physica Scripta}\ }\textbf {\bibinfo {volume}
  {T152}},\ \bibinfo {pages} {014008} (\bibinfo {year} {2013})}\BibitemShut
  {NoStop}%
\bibitem [{\citenamefont {Margueron}\ \emph {et~al.}(2018)\citenamefont
  {Margueron}, \citenamefont {Hoffmann~Casali},\ and\ \citenamefont
  {Gulminelli}}]{Margueron2018}%
  \BibitemOpen
  \bibfield  {author} {\bibinfo {author} {\bibfnamefont {J.}~\bibnamefont
  {Margueron}}, \bibinfo {author} {\bibfnamefont {R.}~\bibnamefont
  {Hoffmann~Casali}}, \ and\ \bibinfo {author} {\bibfnamefont {F.}~\bibnamefont
  {Gulminelli}},\ }\href {\doibase 10.1103/PhysRevC.97.025805} {\bibfield
  {journal} {\bibinfo  {journal} {Phys. Rev. C}\ }\textbf {\bibinfo {volume}
  {97}},\ \bibinfo {pages} {025805} (\bibinfo {year} {2018})}\BibitemShut
  {NoStop}%
\bibitem [{\citenamefont {Demorest}\ \emph {et~al.}(2010)\citenamefont
  {Demorest}, \citenamefont {Pennucci}, \citenamefont {Ransom}, \citenamefont
  {Roberts},\ and\ \citenamefont {Hessels}}]{Demorest2010}%
  \BibitemOpen
  \bibfield  {author} {\bibinfo {author} {\bibfnamefont {P.}~\bibnamefont
  {Demorest}}, \bibinfo {author} {\bibfnamefont {T.}~\bibnamefont {Pennucci}},
  \bibinfo {author} {\bibfnamefont {S.}~\bibnamefont {Ransom}}, \bibinfo
  {author} {\bibfnamefont {M.}~\bibnamefont {Roberts}}, \ and\ \bibinfo
  {author} {\bibfnamefont {J.}~\bibnamefont {Hessels}},\ }\href {\doibase
  10.1038/nature09466} {\bibfield  {journal} {\bibinfo  {journal} {Nature}\
  }\textbf {\bibinfo {volume} {467}},\ \bibinfo {pages} {1081} (\bibinfo {year}
  {2010})},\ \Eprint {http://arxiv.org/abs/1010.5788} {arXiv:1010.5788
  [astro-ph.HE]} \BibitemShut {NoStop}%
%%CITATION = ARXIV:1010.5788;%%
\bibitem [{\citenamefont {Antoniadis}\ \emph {et~al.}(2013)\citenamefont
  {Antoniadis}, \citenamefont {Freire}, \citenamefont {Wex}, \citenamefont
  {Tauris}, \citenamefont {Lynch}, \citenamefont {van Kerkwijk}, \citenamefont
  {Kramer}, \citenamefont {Bassa}, \citenamefont {Dhillon}, \citenamefont
  {Driebe}, \citenamefont {Hessels}, \citenamefont {Kaspi}, \citenamefont
  {Kondratiev}, \citenamefont {Langer}, \citenamefont {Marsh}, \citenamefont
  {McLaughlin}, \citenamefont {Pennucci}, \citenamefont {Ransom}, \citenamefont
  {Stairs}, \citenamefont {van Leeuwen}, \citenamefont {Verbiest},\ and\
  \citenamefont {Whelan}}]{Antoniadis2013}%
  \BibitemOpen
  \bibfield  {author} {\bibinfo {author} {\bibfnamefont {J.}~\bibnamefont
  {Antoniadis}}, \bibinfo {author} {\bibfnamefont {P.~C.~C.}\ \bibnamefont
  {Freire}}, \bibinfo {author} {\bibfnamefont {N.}~\bibnamefont {Wex}},
  \bibinfo {author} {\bibfnamefont {T.~M.}\ \bibnamefont {Tauris}}, \bibinfo
  {author} {\bibfnamefont {R.~S.}\ \bibnamefont {Lynch}}, \bibinfo {author}
  {\bibfnamefont {M.~H.}\ \bibnamefont {van Kerkwijk}}, \bibinfo {author}
  {\bibfnamefont {M.}~\bibnamefont {Kramer}}, \bibinfo {author} {\bibfnamefont
  {C.}~\bibnamefont {Bassa}}, \bibinfo {author} {\bibfnamefont {V.~S.}\
  \bibnamefont {Dhillon}}, \bibinfo {author} {\bibfnamefont {T.}~\bibnamefont
  {Driebe}}, \bibinfo {author} {\bibfnamefont {J.~W.~T.}\ \bibnamefont
  {Hessels}}, \bibinfo {author} {\bibfnamefont {V.~M.}\ \bibnamefont {Kaspi}},
  \bibinfo {author} {\bibfnamefont {V.~I.}\ \bibnamefont {Kondratiev}},
  \bibinfo {author} {\bibfnamefont {N.}~\bibnamefont {Langer}}, \bibinfo
  {author} {\bibfnamefont {T.~R.}\ \bibnamefont {Marsh}}, \bibinfo {author}
  {\bibfnamefont {M.~A.}\ \bibnamefont {McLaughlin}}, \bibinfo {author}
  {\bibfnamefont {T.~T.}\ \bibnamefont {Pennucci}}, \bibinfo {author}
  {\bibfnamefont {S.~M.}\ \bibnamefont {Ransom}}, \bibinfo {author}
  {\bibfnamefont {I.~H.}\ \bibnamefont {Stairs}}, \bibinfo {author}
  {\bibfnamefont {J.}~\bibnamefont {van Leeuwen}}, \bibinfo {author}
  {\bibfnamefont {J.~P.~W.}\ \bibnamefont {Verbiest}}, \ and\ \bibinfo {author}
  {\bibfnamefont {D.~G.}\ \bibnamefont {Whelan}},\ }\href {\doibase
  10.1126/science.1233232} {\bibfield  {journal} {\bibinfo  {journal}
  {Science}\ }\textbf {\bibinfo {volume} {340}} (\bibinfo {year} {2013}),\
  10.1126/science.1233232}\BibitemShut {NoStop}%
\bibitem [{\citenamefont {Margalit}\ and\ \citenamefont
  {Metzger}(2017)}]{Margalit2017}%
  \BibitemOpen
  \bibfield  {author} {\bibinfo {author} {\bibfnamefont {B.}~\bibnamefont
  {Margalit}}\ and\ \bibinfo {author} {\bibfnamefont {B.~D.}\ \bibnamefont
  {Metzger}},\ }\href {\doibase 10.3847/2041-8213/aa991c} {\bibfield  {journal}
  {\bibinfo  {journal} {The Astrophysical Journal}\ }\textbf {\bibinfo {volume}
  {850}},\ \bibinfo {pages} {L19} (\bibinfo {year} {2017})}\BibitemShut
  {NoStop}%
\bibitem [{\citenamefont {{Baym}}\ \emph {et~al.}(2019)\citenamefont {{Baym}},
  \citenamefont {{Furusawa}}, \citenamefont {{Hatsuda}}, \citenamefont
  {{Kojo}},\ and\ \citenamefont {{Togashi}}}]{Baym2019}%
  \BibitemOpen
  \bibfield  {author} {\bibinfo {author} {\bibfnamefont {G.}~\bibnamefont
  {{Baym}}}, \bibinfo {author} {\bibfnamefont {S.}~\bibnamefont {{Furusawa}}},
  \bibinfo {author} {\bibfnamefont {T.}~\bibnamefont {{Hatsuda}}}, \bibinfo
  {author} {\bibfnamefont {T.}~\bibnamefont {{Kojo}}}, \ and\ \bibinfo {author}
  {\bibfnamefont {H.}~\bibnamefont {{Togashi}}},\ }\href@noop {} {\bibfield
  {journal} {\bibinfo  {journal} {arXiv e-prints}\ } (\bibinfo {year}
  {2019})},\ \Eprint {http://arxiv.org/abs/1903.08963} {arXiv:1903.08963
  [astro-ph.HE]} \BibitemShut {NoStop}%
\bibitem [{\citenamefont {Li}\ \emph {et~al.}(2019{\natexlab{b}})\citenamefont
  {Li}, \citenamefont {Krastev}, \citenamefont {Wen}, \citenamefont {Xie},\
  and\ \citenamefont {Zhang}}]{Baoan2019}%
  \BibitemOpen
  \bibfield  {author} {\bibinfo {author} {\bibfnamefont {B.-A.}\ \bibnamefont
  {Li}}, \bibinfo {author} {\bibfnamefont {P.~G.}\ \bibnamefont {Krastev}},
  \bibinfo {author} {\bibfnamefont {D.-H.}\ \bibnamefont {Wen}}, \bibinfo
  {author} {\bibfnamefont {W.-J.}\ \bibnamefont {Xie}}, \ and\ \bibinfo
  {author} {\bibfnamefont {N.-B.}\ \bibnamefont {Zhang}},\ }\href {\doibase
  10.1063/1.5117808} {\bibfield  {journal} {\bibinfo  {journal} {AIP Conference
  Proceedings}\ }\textbf {\bibinfo {volume} {2127}},\ \bibinfo {pages} {020018}
  (\bibinfo {year} {2019}{\natexlab{b}})},\ \Eprint
  {http://arxiv.org/abs/\url{https://aip.scitation.org/doi/pdf/10.1063/1.5117808}}
  {\url{https://aip.scitation.org/doi/pdf/10.1063/1.5117808}} \BibitemShut
  {NoStop}%
\bibitem [{\citenamefont {Shibata}\ \emph {et~al.}(2017)\citenamefont
  {Shibata}, \citenamefont {Fujibayashi}, \citenamefont {Hotokezaka},
  \citenamefont {Kiuchi}, \citenamefont {Kyutoku}, \citenamefont {Sekiguchi},\
  and\ \citenamefont {Tanaka}}]{Shibata2017}%
  \BibitemOpen
  \bibfield  {author} {\bibinfo {author} {\bibfnamefont {M.}~\bibnamefont
  {Shibata}}, \bibinfo {author} {\bibfnamefont {S.}~\bibnamefont
  {Fujibayashi}}, \bibinfo {author} {\bibfnamefont {K.}~\bibnamefont
  {Hotokezaka}}, \bibinfo {author} {\bibfnamefont {K.}~\bibnamefont {Kiuchi}},
  \bibinfo {author} {\bibfnamefont {K.}~\bibnamefont {Kyutoku}}, \bibinfo
  {author} {\bibfnamefont {Y.}~\bibnamefont {Sekiguchi}}, \ and\ \bibinfo
  {author} {\bibfnamefont {M.}~\bibnamefont {Tanaka}},\ }\href {\doibase
  10.1103/PhysRevD.96.123012} {\bibfield  {journal} {\bibinfo  {journal} {Phys.
  Rev. D}\ }\textbf {\bibinfo {volume} {96}},\ \bibinfo {pages} {123012}
  (\bibinfo {year} {2017})}\BibitemShut {NoStop}%
\bibitem [{\citenamefont {Rezzolla}\ \emph {et~al.}(2018)\citenamefont
  {Rezzolla}, \citenamefont {Most},\ and\ \citenamefont {Weih}}]{Rezzolla2018}%
  \BibitemOpen
  \bibfield  {author} {\bibinfo {author} {\bibfnamefont {L.}~\bibnamefont
  {Rezzolla}}, \bibinfo {author} {\bibfnamefont {E.~R.}\ \bibnamefont {Most}},
  \ and\ \bibinfo {author} {\bibfnamefont {L.~R.}\ \bibnamefont {Weih}},\
  }\href {\doibase 10.3847/2041-8213/aaa401} {\bibfield  {journal} {\bibinfo
  {journal} {The Astrophysical Journal}\ }\textbf {\bibinfo {volume} {852}},\
  \bibinfo {pages} {L25} (\bibinfo {year} {2018})}\BibitemShut {NoStop}%
\bibitem [{\citenamefont {Ruiz}\ \emph {et~al.}(2018)\citenamefont {Ruiz},
  \citenamefont {Shapiro},\ and\ \citenamefont {Tsokaros}}]{Ruiz2018}%
  \BibitemOpen
  \bibfield  {author} {\bibinfo {author} {\bibfnamefont {M.}~\bibnamefont
  {Ruiz}}, \bibinfo {author} {\bibfnamefont {S.~L.}\ \bibnamefont {Shapiro}}, \
  and\ \bibinfo {author} {\bibfnamefont {A.}~\bibnamefont {Tsokaros}},\ }\href
  {\doibase 10.1103/PhysRevD.97.021501} {\bibfield  {journal} {\bibinfo
  {journal} {Phys. Rev. D}\ }\textbf {\bibinfo {volume} {97}},\ \bibinfo
  {pages} {021501} (\bibinfo {year} {2018})}\BibitemShut {NoStop}%
\bibitem [{\citenamefont {Zhou}\ \emph {et~al.}(2018)\citenamefont {Zhou},
  \citenamefont {Zhou},\ and\ \citenamefont {Li}}]{Zhou2018}%
  \BibitemOpen
  \bibfield  {author} {\bibinfo {author} {\bibfnamefont {E.-P.}\ \bibnamefont
  {Zhou}}, \bibinfo {author} {\bibfnamefont {X.}~\bibnamefont {Zhou}}, \ and\
  \bibinfo {author} {\bibfnamefont {A.}~\bibnamefont {Li}},\ }\href {\doibase
  10.1103/PhysRevD.97.083015} {\bibfield  {journal} {\bibinfo  {journal} {Phys.
  Rev. D}\ }\textbf {\bibinfo {volume} {97}},\ \bibinfo {pages} {083015}
  (\bibinfo {year} {2018})}\BibitemShut {NoStop}%
\bibitem [{\citenamefont {Lipparini}\ and\ \citenamefont
  {Stringari}(1989)}]{LIPPARINI1989}%
  \BibitemOpen
  \bibfield  {author} {\bibinfo {author} {\bibfnamefont {E.}~\bibnamefont
  {Lipparini}}\ and\ \bibinfo {author} {\bibfnamefont {S.}~\bibnamefont
  {Stringari}},\ }\href {\doibase https://doi.org/10.1016/0370-1573(89)90029-X}
  {\bibfield  {journal} {\bibinfo  {journal} {Physics Reports}\ }\textbf
  {\bibinfo {volume} {175}},\ \bibinfo {pages} {103 } (\bibinfo {year}
  {1989})}\BibitemShut {NoStop}%
\bibitem [{\citenamefont {Tews}\ \emph
  {et~al.}(2018{\natexlab{a}})\citenamefont {Tews}, \citenamefont {Margueron},\
  and\ \citenamefont {Reddy}}]{Tews2018_2}%
  \BibitemOpen
  \bibfield  {author} {\bibinfo {author} {\bibfnamefont {I.}~\bibnamefont
  {Tews}}, \bibinfo {author} {\bibfnamefont {J.}~\bibnamefont {Margueron}}, \
  and\ \bibinfo {author} {\bibfnamefont {S.}~\bibnamefont {Reddy}},\ }\href
  {\doibase 10.1103/PhysRevC.98.045804} {\bibfield  {journal} {\bibinfo
  {journal} {Phys. Rev. C}\ }\textbf {\bibinfo {volume} {98}},\ \bibinfo
  {pages} {045804} (\bibinfo {year} {2018}{\natexlab{a}})}\BibitemShut
  {NoStop}%
\bibitem [{\citenamefont {Guven}\ \emph {et~al.}(2020)\citenamefont {Guven},
  \citenamefont {Bozkurt}, \citenamefont {Khan},\ and\ \citenamefont
  {Margueron}}]{Guven2020}%
  \BibitemOpen
  \bibfield  {author} {\bibinfo {author} {\bibfnamefont {H.}~\bibnamefont
  {Guven}}, \bibinfo {author} {\bibfnamefont {K.}~\bibnamefont {Bozkurt}},
  \bibinfo {author} {\bibfnamefont {E.}~\bibnamefont {Khan}}, \ and\ \bibinfo
  {author} {\bibfnamefont {J.}~\bibnamefont {Margueron}},\ }\href@noop {} {\
  (\bibinfo {year} {2020})},\ \Eprint {http://arxiv.org/abs/2001.10259}
  {arXiv:2001.10259 [nucl-th]} \BibitemShut {NoStop}%
%%CITATION = ARXIV:2001.10259;%%
\bibitem [{\citenamefont {Landau}\ and\ \citenamefont
  {Lifshitz}(1987)}]{Fluid}%
  \BibitemOpen
  \bibfield  {author} {\bibinfo {author} {\bibfnamefont {L.~D.}\ \bibnamefont
  {Landau}}\ and\ \bibinfo {author} {\bibfnamefont {E.~M.}\ \bibnamefont
  {Lifshitz}},\ }\href@noop {} {\emph {\bibinfo {title} {Fluid Mechanics}}}\
  (\bibinfo  {publisher} {Pergamonm Oxford},\ \bibinfo {year} {1987})\ pp.\
  \bibinfo {pages} {251--254}\BibitemShut {NoStop}%
\bibitem [{\citenamefont {Lattimer}\ and\ \citenamefont
  {Prakash}(2016)}]{Lattimer2015}%
  \BibitemOpen
  \bibfield  {author} {\bibinfo {author} {\bibfnamefont {J.~M.}\ \bibnamefont
  {Lattimer}}\ and\ \bibinfo {author} {\bibfnamefont {M.}~\bibnamefont
  {Prakash}},\ }\href {\doibase 10.1016/j.physrep.2015.12.005} {\bibfield
  {journal} {\bibinfo  {journal} {Phys. Rept.}\ }\textbf {\bibinfo {volume}
  {621}},\ \bibinfo {pages} {127} (\bibinfo {year} {2016})},\ \Eprint
  {http://arxiv.org/abs/1512.07820} {arXiv:1512.07820 [astro-ph.SR]}
  \BibitemShut {NoStop}%
%%CITATION = ARXIV:1512.07820;%%
\bibitem [{\citenamefont {Yakovlev}\ \emph {et~al.}(2013)\citenamefont
  {Yakovlev}, \citenamefont {Haensel}, \citenamefont {Baym},\ and\
  \citenamefont {Pethick}}]{Yakovlev2013}%
  \BibitemOpen
  \bibfield  {author} {\bibinfo {author} {\bibfnamefont {D.~G.}\ \bibnamefont
  {Yakovlev}}, \bibinfo {author} {\bibfnamefont {P.}~\bibnamefont {Haensel}},
  \bibinfo {author} {\bibfnamefont {G.}~\bibnamefont {Baym}}, \ and\ \bibinfo
  {author} {\bibfnamefont {C.}~\bibnamefont {Pethick}},\ }\href {\doibase
  10.3367/ufne.0183.201303f.0307} {\bibfield  {journal} {\bibinfo  {journal}
  {Physics-Uspekhi}\ }\textbf {\bibinfo {volume} {56}},\ \bibinfo {pages} {289}
  (\bibinfo {year} {2013})}\BibitemShut {NoStop}%
\bibitem [{\citenamefont {Haensel}\ \emph {et~al.}(2007)\citenamefont
  {Haensel}, \citenamefont {Potekhin},\ and\ \citenamefont
  {Yakovlev}}]{NeutronStarsI}%
  \BibitemOpen
  \bibfield  {author} {\bibinfo {author} {\bibfnamefont {P.}~\bibnamefont
  {Haensel}}, \bibinfo {author} {\bibfnamefont {A.~Y.}\ \bibnamefont
  {Potekhin}}, \ and\ \bibinfo {author} {\bibfnamefont {D.~G.}\ \bibnamefont
  {Yakovlev}},\ }\href@noop {} {\emph {\bibinfo {title} {Neutron Stars I}}}\
  (\bibinfo  {publisher} {Springer},\ \bibinfo {year} {2007})\BibitemShut
  {NoStop}%
\bibitem [{\citenamefont {Nakazato}\ \emph {et~al.}(2011)\citenamefont
  {Nakazato}, \citenamefont {Iida},\ and\ \citenamefont {Oyamatsu}}]{pasta1}%
  \BibitemOpen
  \bibfield  {author} {\bibinfo {author} {\bibfnamefont {K.}~\bibnamefont
  {Nakazato}}, \bibinfo {author} {\bibfnamefont {K.}~\bibnamefont {Iida}}, \
  and\ \bibinfo {author} {\bibfnamefont {K.}~\bibnamefont {Oyamatsu}},\ }\href
  {\doibase 10.1103/PhysRevC.83.065811} {\bibfield  {journal} {\bibinfo
  {journal} {Phys. Rev. C}\ }\textbf {\bibinfo {volume} {83}},\ \bibinfo
  {pages} {065811} (\bibinfo {year} {2011})}\BibitemShut {NoStop}%
\bibitem [{\citenamefont {S\'ebille}\ \emph {et~al.}(2011)\citenamefont
  {S\'ebille}, \citenamefont {de~la Mota},\ and\ \citenamefont
  {Figerou}}]{pasta2}%
  \BibitemOpen
  \bibfield  {author} {\bibinfo {author} {\bibfnamefont {F.}~\bibnamefont
  {S\'ebille}}, \bibinfo {author} {\bibfnamefont {V.}~\bibnamefont {de~la
  Mota}}, \ and\ \bibinfo {author} {\bibfnamefont {S.}~\bibnamefont
  {Figerou}},\ }\href {\doibase 10.1103/PhysRevC.84.055801} {\bibfield
  {journal} {\bibinfo  {journal} {Phys. Rev. C}\ }\textbf {\bibinfo {volume}
  {84}},\ \bibinfo {pages} {055801} (\bibinfo {year} {2011})}\BibitemShut
  {NoStop}%
\bibitem [{\citenamefont {Vantournhout}\ \emph {et~al.}(2011)\citenamefont
  {Vantournhout}, \citenamefont {Neff}, \citenamefont {Feldmeier},
  \citenamefont {Jachowicz},\ and\ \citenamefont {Ryckebusch}}]{pasta3}%
  \BibitemOpen
  \bibfield  {author} {\bibinfo {author} {\bibfnamefont {K.}~\bibnamefont
  {Vantournhout}}, \bibinfo {author} {\bibfnamefont {T.}~\bibnamefont {Neff}},
  \bibinfo {author} {\bibfnamefont {H.}~\bibnamefont {Feldmeier}}, \bibinfo
  {author} {\bibfnamefont {N.}~\bibnamefont {Jachowicz}}, \ and\ \bibinfo
  {author} {\bibfnamefont {J.}~\bibnamefont {Ryckebusch}},\ }\href {\doibase
  10.1016/j.ppnp.2011.01.019} {\bibfield  {journal} {\bibinfo  {journal} {Prog.
  Part. Nucl. Phys.}\ }\textbf {\bibinfo {volume} {66}},\ \bibinfo {pages}
  {271} (\bibinfo {year} {2011})},\ \Eprint {http://arxiv.org/abs/1011.2928}
  {arXiv:1011.2928 [nucl-th]} \BibitemShut {NoStop}%
\bibitem [{\citenamefont {Pais}\ and\ \citenamefont {Stone}(2012)}]{pasta4}%
  \BibitemOpen
  \bibfield  {author} {\bibinfo {author} {\bibfnamefont {H.}~\bibnamefont
  {Pais}}\ and\ \bibinfo {author} {\bibfnamefont {J.~R.}\ \bibnamefont
  {Stone}},\ }\href {\doibase 10.1103/PhysRevLett.109.151101} {\bibfield
  {journal} {\bibinfo  {journal} {Phys. Rev. Lett.}\ }\textbf {\bibinfo
  {volume} {109}},\ \bibinfo {pages} {151101} (\bibinfo {year}
  {2012})}\BibitemShut {NoStop}%
\bibitem [{\citenamefont {Gupta}\ and\ \citenamefont
  {Arumugam}(2013)}]{pasta5}%
  \BibitemOpen
  \bibfield  {author} {\bibinfo {author} {\bibfnamefont {N.}~\bibnamefont
  {Gupta}}\ and\ \bibinfo {author} {\bibfnamefont {P.}~\bibnamefont
  {Arumugam}},\ }\href {\doibase 10.1103/PhysRevC.87.028801} {\bibfield
  {journal} {\bibinfo  {journal} {Phys. Rev. C}\ }\textbf {\bibinfo {volume}
  {87}},\ \bibinfo {pages} {028801} (\bibinfo {year} {2013})}\BibitemShut
  {NoStop}%
\bibitem [{\citenamefont {Okamoto}\ \emph {et~al.}(2013)\citenamefont
  {Okamoto}, \citenamefont {Maruyama}, \citenamefont {Yabana},\ and\
  \citenamefont {Tatsumi}}]{pasta6}%
  \BibitemOpen
  \bibfield  {author} {\bibinfo {author} {\bibfnamefont {M.}~\bibnamefont
  {Okamoto}}, \bibinfo {author} {\bibfnamefont {T.}~\bibnamefont {Maruyama}},
  \bibinfo {author} {\bibfnamefont {K.}~\bibnamefont {Yabana}}, \ and\ \bibinfo
  {author} {\bibfnamefont {T.}~\bibnamefont {Tatsumi}},\ }\href {\doibase
  10.1103/PhysRevC.88.025801} {\bibfield  {journal} {\bibinfo  {journal} {Phys.
  Rev. C}\ }\textbf {\bibinfo {volume} {88}},\ \bibinfo {pages} {025801}
  (\bibinfo {year} {2013})}\BibitemShut {NoStop}%
\bibitem [{\citenamefont {Alcain}\ \emph {et~al.}(2014)\citenamefont {Alcain},
  \citenamefont {Gim\'enez~Molinelli}, \citenamefont {Nichols},\ and\
  \citenamefont {Dorso}}]{pasta7}%
  \BibitemOpen
  \bibfield  {author} {\bibinfo {author} {\bibfnamefont {P.~N.}\ \bibnamefont
  {Alcain}}, \bibinfo {author} {\bibfnamefont {P.~A.}\ \bibnamefont
  {Gim\'enez~Molinelli}}, \bibinfo {author} {\bibfnamefont {J.~I.}\
  \bibnamefont {Nichols}}, \ and\ \bibinfo {author} {\bibfnamefont {C.~O.}\
  \bibnamefont {Dorso}},\ }\href {\doibase 10.1103/PhysRevC.89.055801}
  {\bibfield  {journal} {\bibinfo  {journal} {Phys. Rev. C}\ }\textbf {\bibinfo
  {volume} {89}},\ \bibinfo {pages} {055801} (\bibinfo {year}
  {2014})}\BibitemShut {NoStop}%
\bibitem [{\citenamefont {Grill}\ \emph {et~al.}(2014)\citenamefont {Grill},
  \citenamefont {Pais}, \citenamefont {Provid\^encia}, \citenamefont
  {Vida\~na},\ and\ \citenamefont {Avancini}}]{pasta8}%
  \BibitemOpen
  \bibfield  {author} {\bibinfo {author} {\bibfnamefont {F.}~\bibnamefont
  {Grill}}, \bibinfo {author} {\bibfnamefont {H.}~\bibnamefont {Pais}},
  \bibinfo {author} {\bibfnamefont {C.~m.~c.}\ \bibnamefont {Provid\^encia}},
  \bibinfo {author} {\bibfnamefont {I.}~\bibnamefont {Vida\~na}}, \ and\
  \bibinfo {author} {\bibfnamefont {S.~S.}\ \bibnamefont {Avancini}},\ }\href
  {\doibase 10.1103/PhysRevC.90.045803} {\bibfield  {journal} {\bibinfo
  {journal} {Phys. Rev. C}\ }\textbf {\bibinfo {volume} {90}},\ \bibinfo
  {pages} {045803} (\bibinfo {year} {2014})}\BibitemShut {NoStop}%
\bibitem [{\citenamefont {Martin}\ and\ \citenamefont {Urban}(2015)}]{pasta9}%
  \BibitemOpen
  \bibfield  {author} {\bibinfo {author} {\bibfnamefont {N.}~\bibnamefont
  {Martin}}\ and\ \bibinfo {author} {\bibfnamefont {M.}~\bibnamefont {Urban}},\
  }\href {\doibase 10.1103/PhysRevC.92.015803} {\bibfield  {journal} {\bibinfo
  {journal} {Phys. Rev. C}\ }\textbf {\bibinfo {volume} {92}},\ \bibinfo
  {pages} {015803} (\bibinfo {year} {2015})}\BibitemShut {NoStop}%
\bibitem [{\citenamefont {Molinelli}\ and\ \citenamefont
  {Dorso}(2015)}]{pasta10}%
  \BibitemOpen
  \bibfield  {author} {\bibinfo {author} {\bibfnamefont {P.~G.}\ \bibnamefont
  {Molinelli}}\ and\ \bibinfo {author} {\bibfnamefont {C.}~\bibnamefont
  {Dorso}},\ }\href {\doibase https://doi.org/10.1016/j.nuclphysa.2014.11.005}
  {\bibfield  {journal} {\bibinfo  {journal} {Nuclear Physics A}\ }\textbf
  {\bibinfo {volume} {933}},\ \bibinfo {pages} {306 } (\bibinfo {year}
  {2015})}\BibitemShut {NoStop}%
\bibitem [{\citenamefont {Sagert}\ \emph {et~al.}(2016)\citenamefont {Sagert},
  \citenamefont {Fann}, \citenamefont {Fattoyev}, \citenamefont {Postnikov},\
  and\ \citenamefont {Horowitz}}]{pasta11}%
  \BibitemOpen
  \bibfield  {author} {\bibinfo {author} {\bibfnamefont {I.}~\bibnamefont
  {Sagert}}, \bibinfo {author} {\bibfnamefont {G.~I.}\ \bibnamefont {Fann}},
  \bibinfo {author} {\bibfnamefont {F.~J.}\ \bibnamefont {Fattoyev}}, \bibinfo
  {author} {\bibfnamefont {S.}~\bibnamefont {Postnikov}}, \ and\ \bibinfo
  {author} {\bibfnamefont {C.~J.}\ \bibnamefont {Horowitz}},\ }\href {\doibase
  10.1103/PhysRevC.93.055801} {\bibfield  {journal} {\bibinfo  {journal} {Phys.
  Rev. C}\ }\textbf {\bibinfo {volume} {93}},\ \bibinfo {pages} {055801}
  (\bibinfo {year} {2016})}\BibitemShut {NoStop}%
\bibitem [{\citenamefont {Nandi}\ and\ \citenamefont
  {Schramm}(2016)}]{pasta12}%
  \BibitemOpen
  \bibfield  {author} {\bibinfo {author} {\bibfnamefont {R.}~\bibnamefont
  {Nandi}}\ and\ \bibinfo {author} {\bibfnamefont {S.}~\bibnamefont
  {Schramm}},\ }\href {\doibase 10.1103/PhysRevC.94.025806} {\bibfield
  {journal} {\bibinfo  {journal} {Phys. Rev. C}\ }\textbf {\bibinfo {volume}
  {94}},\ \bibinfo {pages} {025806} (\bibinfo {year} {2016})}\BibitemShut
  {NoStop}%
\bibitem [{\citenamefont {Kubis}\ and\ \citenamefont
  {W\'ojcik}(2016)}]{pasta13}%
  \BibitemOpen
  \bibfield  {author} {\bibinfo {author} {\bibfnamefont {S.}~\bibnamefont
  {Kubis}}\ and\ \bibinfo {author} {\bibfnamefont {W.}~\bibnamefont
  {W\'ojcik}},\ }\href {\doibase 10.1103/PhysRevC.94.065805} {\bibfield
  {journal} {\bibinfo  {journal} {Phys. Rev. C}\ }\textbf {\bibinfo {volume}
  {94}},\ \bibinfo {pages} {065805} (\bibinfo {year} {2016})}\BibitemShut
  {NoStop}%
\bibitem [{\citenamefont {Grams}\ \emph {et~al.}(2017)\citenamefont {Grams},
  \citenamefont {Santos}, \citenamefont {Panda}, \citenamefont
  {Provid\^encia},\ and\ \citenamefont {Menezes}}]{pasta14}%
  \BibitemOpen
  \bibfield  {author} {\bibinfo {author} {\bibfnamefont {G.}~\bibnamefont
  {Grams}}, \bibinfo {author} {\bibfnamefont {A.~M.}\ \bibnamefont {Santos}},
  \bibinfo {author} {\bibfnamefont {P.~K.}\ \bibnamefont {Panda}}, \bibinfo
  {author} {\bibfnamefont {C.~m.~c.}\ \bibnamefont {Provid\^encia}}, \ and\
  \bibinfo {author} {\bibfnamefont {D.~P.}\ \bibnamefont {Menezes}},\ }\href
  {\doibase 10.1103/PhysRevC.95.055807} {\bibfield  {journal} {\bibinfo
  {journal} {Phys. Rev. C}\ }\textbf {\bibinfo {volume} {95}},\ \bibinfo
  {pages} {055807} (\bibinfo {year} {2017})}\BibitemShut {NoStop}%
\bibitem [{\citenamefont {Kycia}\ \emph {et~al.}(2017)\citenamefont {Kycia},
  \citenamefont {Kubis},\ and\ \citenamefont {W\'ojcik}}]{pasta15}%
  \BibitemOpen
  \bibfield  {author} {\bibinfo {author} {\bibfnamefont {R.~A.}\ \bibnamefont
  {Kycia}}, \bibinfo {author} {\bibfnamefont {S.}~\bibnamefont {Kubis}}, \ and\
  \bibinfo {author} {\bibfnamefont {W.}~\bibnamefont {W\'ojcik}},\ }\href
  {\doibase 10.1103/PhysRevC.96.025803} {\bibfield  {journal} {\bibinfo
  {journal} {Phys. Rev. C}\ }\textbf {\bibinfo {volume} {96}},\ \bibinfo
  {pages} {025803} (\bibinfo {year} {2017})}\BibitemShut {NoStop}%
\bibitem [{\citenamefont {Schuetrumpf}\ \emph {et~al.}(2019)\citenamefont
  {Schuetrumpf}, \citenamefont {Mart\'{\i}nez-Pinedo}, \citenamefont
  {Afibuzzaman},\ and\ \citenamefont {Aktulga}}]{pasta16}%
  \BibitemOpen
  \bibfield  {author} {\bibinfo {author} {\bibfnamefont {B.}~\bibnamefont
  {Schuetrumpf}}, \bibinfo {author} {\bibfnamefont {G.}~\bibnamefont
  {Mart\'{\i}nez-Pinedo}}, \bibinfo {author} {\bibfnamefont {M.}~\bibnamefont
  {Afibuzzaman}}, \ and\ \bibinfo {author} {\bibfnamefont {H.~M.}\ \bibnamefont
  {Aktulga}},\ }\href {\doibase 10.1103/PhysRevC.100.045806} {\bibfield
  {journal} {\bibinfo  {journal} {Phys. Rev. C}\ }\textbf {\bibinfo {volume}
  {100}},\ \bibinfo {pages} {045806} (\bibinfo {year} {2019})}\BibitemShut
  {NoStop}%
\bibitem [{\citenamefont {Pearson}\ \emph {et~al.}(2020)\citenamefont
  {Pearson}, \citenamefont {Chamel},\ and\ \citenamefont {Potekhin}}]{pasta17}%
  \BibitemOpen
  \bibfield  {author} {\bibinfo {author} {\bibfnamefont {J.~M.}\ \bibnamefont
  {Pearson}}, \bibinfo {author} {\bibfnamefont {N.}~\bibnamefont {Chamel}}, \
  and\ \bibinfo {author} {\bibfnamefont {A.~Y.}\ \bibnamefont {Potekhin}},\
  }\href {\doibase 10.1103/PhysRevC.101.015802} {\bibfield  {journal} {\bibinfo
   {journal} {Phys. Rev. C}\ }\textbf {\bibinfo {volume} {101}},\ \bibinfo
  {pages} {015802} (\bibinfo {year} {2020})}\BibitemShut {NoStop}%
\bibitem [{\citenamefont {Barros}\ \emph {et~al.}(2020)\citenamefont {Barros},
  \citenamefont {Menezes},\ and\ \citenamefont {Gulminelli}}]{pasta18}%
  \BibitemOpen
  \bibfield  {author} {\bibinfo {author} {\bibfnamefont {C.~C.}\ \bibnamefont
  {Barros}}, \bibinfo {author} {\bibfnamefont {D.~P.}\ \bibnamefont {Menezes}},
  \ and\ \bibinfo {author} {\bibfnamefont {F.}~\bibnamefont {Gulminelli}},\
  }\href {\doibase 10.1103/PhysRevC.101.035211} {\bibfield  {journal} {\bibinfo
   {journal} {Phys. Rev. C}\ }\textbf {\bibinfo {volume} {101}},\ \bibinfo
  {pages} {035211} (\bibinfo {year} {2020})}\BibitemShut {NoStop}%
\bibitem [{\citenamefont {Horowitz}\ \emph
  {et~al.}(2004{\natexlab{a}})\citenamefont {Horowitz}, \citenamefont
  {P\'erez-Garc\'{\i}a},\ and\ \citenamefont {Piekarewicz}}]{Horowitz2004}%
  \BibitemOpen
  \bibfield  {author} {\bibinfo {author} {\bibfnamefont {C.~J.}\ \bibnamefont
  {Horowitz}}, \bibinfo {author} {\bibfnamefont {M.~A.}\ \bibnamefont
  {P\'erez-Garc\'{\i}a}}, \ and\ \bibinfo {author} {\bibfnamefont
  {J.}~\bibnamefont {Piekarewicz}},\ }\href {\doibase
  10.1103/PhysRevC.69.045804} {\bibfield  {journal} {\bibinfo  {journal} {Phys.
  Rev. C}\ }\textbf {\bibinfo {volume} {69}},\ \bibinfo {pages} {045804}
  (\bibinfo {year} {2004}{\natexlab{a}})}\BibitemShut {NoStop}%
\bibitem [{\citenamefont {Horowitz}\ \emph
  {et~al.}(2004{\natexlab{b}})\citenamefont {Horowitz}, \citenamefont
  {P\'erez-Garc\'{\i}a}, \citenamefont {Carriere}, \citenamefont {Berry},\ and\
  \citenamefont {Piekarewicz}}]{Horowitz2004_2}%
  \BibitemOpen
  \bibfield  {author} {\bibinfo {author} {\bibfnamefont {C.~J.}\ \bibnamefont
  {Horowitz}}, \bibinfo {author} {\bibfnamefont {M.~A.}\ \bibnamefont
  {P\'erez-Garc\'{\i}a}}, \bibinfo {author} {\bibfnamefont {J.}~\bibnamefont
  {Carriere}}, \bibinfo {author} {\bibfnamefont {D.~K.}\ \bibnamefont {Berry}},
  \ and\ \bibinfo {author} {\bibfnamefont {J.}~\bibnamefont {Piekarewicz}},\
  }\href {\doibase 10.1103/PhysRevC.70.065806} {\bibfield  {journal} {\bibinfo
  {journal} {Phys. Rev. C}\ }\textbf {\bibinfo {volume} {70}},\ \bibinfo
  {pages} {065806} (\bibinfo {year} {2004}{\natexlab{b}})}\BibitemShut
  {NoStop}%
\bibitem [{\citenamefont {Horowitz}\ \emph {et~al.}(2005)\citenamefont
  {Horowitz}, \citenamefont {P\'erez-Garc\'{\i}a}, \citenamefont {Berry},\ and\
  \citenamefont {Piekarewicz}}]{Horowitz2005}%
  \BibitemOpen
  \bibfield  {author} {\bibinfo {author} {\bibfnamefont {C.~J.}\ \bibnamefont
  {Horowitz}}, \bibinfo {author} {\bibfnamefont {M.~A.}\ \bibnamefont
  {P\'erez-Garc\'{\i}a}}, \bibinfo {author} {\bibfnamefont {D.~K.}\
  \bibnamefont {Berry}}, \ and\ \bibinfo {author} {\bibfnamefont
  {J.}~\bibnamefont {Piekarewicz}},\ }\href {\doibase
  10.1103/PhysRevC.72.035801} {\bibfield  {journal} {\bibinfo  {journal} {Phys.
  Rev. C}\ }\textbf {\bibinfo {volume} {72}},\ \bibinfo {pages} {035801}
  (\bibinfo {year} {2005})}\BibitemShut {NoStop}%
\bibitem [{\citenamefont {Watanabe}\ \emph {et~al.}(2003)\citenamefont
  {Watanabe}, \citenamefont {Sato}, \citenamefont {Yasuoka},\ and\
  \citenamefont {Ebisuzaki}}]{Watanabe2003}%
  \BibitemOpen
  \bibfield  {author} {\bibinfo {author} {\bibfnamefont {G.}~\bibnamefont
  {Watanabe}}, \bibinfo {author} {\bibfnamefont {K.}~\bibnamefont {Sato}},
  \bibinfo {author} {\bibfnamefont {K.}~\bibnamefont {Yasuoka}}, \ and\
  \bibinfo {author} {\bibfnamefont {T.}~\bibnamefont {Ebisuzaki}},\ }\href
  {\doibase 10.1103/PhysRevC.68.035806} {\bibfield  {journal} {\bibinfo
  {journal} {Phys. Rev. C}\ }\textbf {\bibinfo {volume} {68}},\ \bibinfo
  {pages} {035806} (\bibinfo {year} {2003})}\BibitemShut {NoStop}%
\bibitem [{\citenamefont {Watanabe}\ \emph {et~al.}(2005)\citenamefont
  {Watanabe}, \citenamefont {Maruyama}, \citenamefont {Sato}, \citenamefont
  {Yasuoka},\ and\ \citenamefont {Ebisuzaki}}]{Watanabe2005}%
  \BibitemOpen
  \bibfield  {author} {\bibinfo {author} {\bibfnamefont {G.}~\bibnamefont
  {Watanabe}}, \bibinfo {author} {\bibfnamefont {T.}~\bibnamefont {Maruyama}},
  \bibinfo {author} {\bibfnamefont {K.}~\bibnamefont {Sato}}, \bibinfo {author}
  {\bibfnamefont {K.}~\bibnamefont {Yasuoka}}, \ and\ \bibinfo {author}
  {\bibfnamefont {T.}~\bibnamefont {Ebisuzaki}},\ }\href {\doibase
  10.1103/PhysRevLett.94.031101} {\bibfield  {journal} {\bibinfo  {journal}
  {Phys. Rev. Lett.}\ }\textbf {\bibinfo {volume} {94}},\ \bibinfo {pages}
  {031101} (\bibinfo {year} {2005})}\BibitemShut {NoStop}%
\bibitem [{\citenamefont {Watanabe}\ \emph {et~al.}(2009)\citenamefont
  {Watanabe}, \citenamefont {Sonoda}, \citenamefont {Maruyama}, \citenamefont
  {Sato}, \citenamefont {Yasuoka},\ and\ \citenamefont
  {Ebisuzaki}}]{Watanabe2009}%
  \BibitemOpen
  \bibfield  {author} {\bibinfo {author} {\bibfnamefont {G.}~\bibnamefont
  {Watanabe}}, \bibinfo {author} {\bibfnamefont {H.}~\bibnamefont {Sonoda}},
  \bibinfo {author} {\bibfnamefont {T.}~\bibnamefont {Maruyama}}, \bibinfo
  {author} {\bibfnamefont {K.}~\bibnamefont {Sato}}, \bibinfo {author}
  {\bibfnamefont {K.}~\bibnamefont {Yasuoka}}, \ and\ \bibinfo {author}
  {\bibfnamefont {T.}~\bibnamefont {Ebisuzaki}},\ }\href {\doibase
  10.1103/PhysRevLett.103.121101} {\bibfield  {journal} {\bibinfo  {journal}
  {Phys. Rev. Lett.}\ }\textbf {\bibinfo {volume} {103}},\ \bibinfo {pages}
  {121101} (\bibinfo {year} {2009})}\BibitemShut {NoStop}%
\bibitem [{\citenamefont {Schneider}\ \emph {et~al.}(2013)\citenamefont
  {Schneider}, \citenamefont {Horowitz}, \citenamefont {Hughto},\ and\
  \citenamefont {Berry}}]{Schneider2013}%
  \BibitemOpen
  \bibfield  {author} {\bibinfo {author} {\bibfnamefont {A.~S.}\ \bibnamefont
  {Schneider}}, \bibinfo {author} {\bibfnamefont {C.~J.}\ \bibnamefont
  {Horowitz}}, \bibinfo {author} {\bibfnamefont {J.}~\bibnamefont {Hughto}}, \
  and\ \bibinfo {author} {\bibfnamefont {D.~K.}\ \bibnamefont {Berry}},\ }\href
  {\doibase 10.1103/PhysRevC.88.065807} {\bibfield  {journal} {\bibinfo
  {journal} {Phys. Rev. C}\ }\textbf {\bibinfo {volume} {88}},\ \bibinfo
  {pages} {065807} (\bibinfo {year} {2013})}\BibitemShut {NoStop}%
\bibitem [{\citenamefont {Horowitz}\ \emph {et~al.}(2015)\citenamefont
  {Horowitz}, \citenamefont {Berry}, \citenamefont {Briggs}, \citenamefont
  {Caplan}, \citenamefont {Cumming},\ and\ \citenamefont
  {Schneider}}]{Horowitz2015}%
  \BibitemOpen
  \bibfield  {author} {\bibinfo {author} {\bibfnamefont {C.~J.}\ \bibnamefont
  {Horowitz}}, \bibinfo {author} {\bibfnamefont {D.~K.}\ \bibnamefont {Berry}},
  \bibinfo {author} {\bibfnamefont {C.~M.}\ \bibnamefont {Briggs}}, \bibinfo
  {author} {\bibfnamefont {M.~E.}\ \bibnamefont {Caplan}}, \bibinfo {author}
  {\bibfnamefont {A.}~\bibnamefont {Cumming}}, \ and\ \bibinfo {author}
  {\bibfnamefont {A.~S.}\ \bibnamefont {Schneider}},\ }\href {\doibase
  10.1103/PhysRevLett.114.031102} {\bibfield  {journal} {\bibinfo  {journal}
  {Phys. Rev. Lett.}\ }\textbf {\bibinfo {volume} {114}},\ \bibinfo {pages}
  {031102} (\bibinfo {year} {2015})}\BibitemShut {NoStop}%
\bibitem [{\citenamefont {Caplan}\ \emph {et~al.}(2015)\citenamefont {Caplan},
  \citenamefont {Schneider}, \citenamefont {Horowitz},\ and\ \citenamefont
  {Berry}}]{Caplan2015}%
  \BibitemOpen
  \bibfield  {author} {\bibinfo {author} {\bibfnamefont {M.~E.}\ \bibnamefont
  {Caplan}}, \bibinfo {author} {\bibfnamefont {A.~S.}\ \bibnamefont
  {Schneider}}, \bibinfo {author} {\bibfnamefont {C.~J.}\ \bibnamefont
  {Horowitz}}, \ and\ \bibinfo {author} {\bibfnamefont {D.~K.}\ \bibnamefont
  {Berry}},\ }\href {\doibase 10.1103/PhysRevC.91.065802} {\bibfield  {journal}
  {\bibinfo  {journal} {Phys. Rev. C}\ }\textbf {\bibinfo {volume} {91}},\
  \bibinfo {pages} {065802} (\bibinfo {year} {2015})}\BibitemShut {NoStop}%
\bibitem [{\citenamefont {Magierski}\ and\ \citenamefont
  {Heenen}(2002)}]{Magierski2002}%
  \BibitemOpen
  \bibfield  {author} {\bibinfo {author} {\bibfnamefont {P.}~\bibnamefont
  {Magierski}}\ and\ \bibinfo {author} {\bibfnamefont {P.-H.}\ \bibnamefont
  {Heenen}},\ }\href {\doibase 10.1103/PhysRevC.65.045804} {\bibfield
  {journal} {\bibinfo  {journal} {Phys. Rev. C}\ }\textbf {\bibinfo {volume}
  {65}},\ \bibinfo {pages} {045804} (\bibinfo {year} {2002})}\BibitemShut
  {NoStop}%
\bibitem [{\citenamefont {Chamel}(2005)}]{Chamel2005}%
  \BibitemOpen
  \bibfield  {author} {\bibinfo {author} {\bibfnamefont {N.}~\bibnamefont
  {Chamel}},\ }\href {\doibase https://doi.org/10.1016/j.nuclphysa.2004.09.011}
  {\bibfield  {journal} {\bibinfo  {journal} {Nuclear Physics A}\ }\textbf
  {\bibinfo {volume} {747}},\ \bibinfo {pages} {109 } (\bibinfo {year}
  {2005})}\BibitemShut {NoStop}%
\bibitem [{\citenamefont {Newton}\ and\ \citenamefont
  {Stone}(2009)}]{Newton2009}%
  \BibitemOpen
  \bibfield  {author} {\bibinfo {author} {\bibfnamefont {W.~G.}\ \bibnamefont
  {Newton}}\ and\ \bibinfo {author} {\bibfnamefont {J.~R.}\ \bibnamefont
  {Stone}},\ }\href {\doibase 10.1103/PhysRevC.79.055801} {\bibfield  {journal}
  {\bibinfo  {journal} {Phys. Rev. C}\ }\textbf {\bibinfo {volume} {79}},\
  \bibinfo {pages} {055801} (\bibinfo {year} {2009})}\BibitemShut {NoStop}%
\bibitem [{\citenamefont {Schuetrumpf}\ and\ \citenamefont
  {Nazarewicz}(2015)}]{Schuetrumpf2015}%
  \BibitemOpen
  \bibfield  {author} {\bibinfo {author} {\bibfnamefont {B.}~\bibnamefont
  {Schuetrumpf}}\ and\ \bibinfo {author} {\bibfnamefont {W.}~\bibnamefont
  {Nazarewicz}},\ }\href {\doibase 10.1103/PhysRevC.92.045806} {\bibfield
  {journal} {\bibinfo  {journal} {Phys. Rev. C}\ }\textbf {\bibinfo {volume}
  {92}},\ \bibinfo {pages} {045806} (\bibinfo {year} {2015})}\BibitemShut
  {NoStop}%
\bibitem [{\citenamefont {Fattoyev}\ \emph {et~al.}(2017)\citenamefont
  {Fattoyev}, \citenamefont {Horowitz},\ and\ \citenamefont
  {Schuetrumpf}}]{Fattoyev2017}%
  \BibitemOpen
  \bibfield  {author} {\bibinfo {author} {\bibfnamefont {F.~J.}\ \bibnamefont
  {Fattoyev}}, \bibinfo {author} {\bibfnamefont {C.~J.}\ \bibnamefont
  {Horowitz}}, \ and\ \bibinfo {author} {\bibfnamefont {B.}~\bibnamefont
  {Schuetrumpf}},\ }\href {\doibase 10.1103/PhysRevC.95.055804} {\bibfield
  {journal} {\bibinfo  {journal} {Phys. Rev. C}\ }\textbf {\bibinfo {volume}
  {95}},\ \bibinfo {pages} {055804} (\bibinfo {year} {2017})}\BibitemShut
  {NoStop}%
\bibitem [{\citenamefont {Ravenhall}\ \emph {et~al.}(1983)\citenamefont
  {Ravenhall}, \citenamefont {Pethick},\ and\ \citenamefont
  {Wilson}}]{Ravenhall1983}%
  \BibitemOpen
  \bibfield  {author} {\bibinfo {author} {\bibfnamefont {D.~G.}\ \bibnamefont
  {Ravenhall}}, \bibinfo {author} {\bibfnamefont {C.~J.}\ \bibnamefont
  {Pethick}}, \ and\ \bibinfo {author} {\bibfnamefont {J.~R.}\ \bibnamefont
  {Wilson}},\ }\href {\doibase 10.1103/PhysRevLett.50.2066} {\bibfield
  {journal} {\bibinfo  {journal} {Phys. Rev. Lett.}\ }\textbf {\bibinfo
  {volume} {50}},\ \bibinfo {pages} {2066} (\bibinfo {year}
  {1983})}\BibitemShut {NoStop}%
\bibitem [{\citenamefont {Hashimoto}\ \emph {et~al.}(1984)\citenamefont
  {Hashimoto}, \citenamefont {Seki},\ and\ \citenamefont
  {Yamada}}]{Hashimoto:1984}%
  \BibitemOpen
  \bibfield  {author} {\bibinfo {author} {\bibfnamefont {M.}~\bibnamefont
  {Hashimoto}}, \bibinfo {author} {\bibfnamefont {H.}~\bibnamefont {Seki}}, \
  and\ \bibinfo {author} {\bibfnamefont {M.}~\bibnamefont {Yamada}},\
  }\href@noop {} {\bibfield  {journal} {\bibinfo  {journal} {Prog. Theor.
  Phys.}\ }\textbf {\bibinfo {volume} {71}},\ \bibinfo {pages} {320} (\bibinfo
  {year} {1984})}\BibitemShut {NoStop}%
\bibitem [{\citenamefont {Oyamatsu}\ \emph {et~al.}(1984)\citenamefont
  {Oyamatsu}, \citenamefont {Hashimoto},\ and\ \citenamefont
  {Yamada}}]{Oyamatsu:1984}%
  \BibitemOpen
  \bibfield  {author} {\bibinfo {author} {\bibfnamefont {K.}~\bibnamefont
  {Oyamatsu}}, \bibinfo {author} {\bibfnamefont {M.-a.}\ \bibnamefont
  {Hashimoto}}, \ and\ \bibinfo {author} {\bibfnamefont {M.}~\bibnamefont
  {Yamada}},\ }\href@noop {} {\bibfield  {journal} {\bibinfo  {journal} {Prog.
  Theor. Phys.}\ }\textbf {\bibinfo {volume} {72}},\ \bibinfo {pages} {373}
  (\bibinfo {year} {1984})}\BibitemShut {NoStop}%
\bibitem [{\citenamefont {Ambartsumyan}\ and\ \citenamefont
  {Saakyan}(1960)}]{Ambartsumyan1960}%
  \BibitemOpen
  \bibfield  {author} {\bibinfo {author} {\bibfnamefont {V.~A.}\ \bibnamefont
  {Ambartsumyan}}\ and\ \bibinfo {author} {\bibfnamefont {G.~S.}\ \bibnamefont
  {Saakyan}},\ }\href@noop {} {\bibfield  {journal} {\bibinfo  {journal}
  {Soviet Astronomy}\ }\textbf {\bibinfo {volume} {4}},\ \bibinfo {pages} {187}
  (\bibinfo {year} {1960})}\BibitemShut {NoStop}%
\bibitem [{\citenamefont {Chatterjee}\ and\ \citenamefont
  {Vida{\~n}a}(2016)}]{Chatterjee2016}%
  \BibitemOpen
  \bibfield  {author} {\bibinfo {author} {\bibfnamefont {D.}~\bibnamefont
  {Chatterjee}}\ and\ \bibinfo {author} {\bibfnamefont {I.}~\bibnamefont
  {Vida{\~n}a}},\ }\href {\doibase 10.1140/epja/i2016-16029-x} {\bibfield
  {journal} {\bibinfo  {journal} {The European Physical Journal A}\ }\textbf
  {\bibinfo {volume} {52}},\ \bibinfo {pages} {29} (\bibinfo {year}
  {2016})}\BibitemShut {NoStop}%
\bibitem [{\citenamefont {Ducoin}\ \emph {et~al.}(2011)\citenamefont {Ducoin},
  \citenamefont {Margueron}, \citenamefont {Provid\^encia},\ and\ \citenamefont
  {Vida\~na}}]{Ducoin2011}%
  \BibitemOpen
  \bibfield  {author} {\bibinfo {author} {\bibfnamefont {C.}~\bibnamefont
  {Ducoin}}, \bibinfo {author} {\bibfnamefont {J.}~\bibnamefont {Margueron}},
  \bibinfo {author} {\bibfnamefont {C.}~\bibnamefont {Provid\^encia}}, \ and\
  \bibinfo {author} {\bibfnamefont {I.}~\bibnamefont {Vida\~na}},\ }\href
  {\doibase 10.1103/PhysRevC.83.045810} {\bibfield  {journal} {\bibinfo
  {journal} {Phys. Rev. C}\ }\textbf {\bibinfo {volume} {83}},\ \bibinfo
  {pages} {045810} (\bibinfo {year} {2011})}\BibitemShut {NoStop}%
\bibitem [{\citenamefont {Baym}\ \emph {et~al.}(1971)\citenamefont {Baym},
  \citenamefont {Pethick},\ and\ \citenamefont {Sutherland}}]{Baym1971}%
  \BibitemOpen
  \bibfield  {author} {\bibinfo {author} {\bibfnamefont {G.}~\bibnamefont
  {Baym}}, \bibinfo {author} {\bibfnamefont {C.}~\bibnamefont {Pethick}}, \
  and\ \bibinfo {author} {\bibfnamefont {P.}~\bibnamefont {Sutherland}},\
  }\href {\doibase 10.1086/151216} {\bibfield  {journal} {\bibinfo  {journal}
  {Astrophys. J.}\ }\textbf {\bibinfo {volume} {170}},\ \bibinfo {pages} {299}
  (\bibinfo {year} {1971})}\BibitemShut {NoStop}%
%%CITATION = ASJOA,170,299;%%
\bibitem [{\citenamefont {Ji}\ \emph {et~al.}(2019)\citenamefont {Ji},
  \citenamefont {Hu}, \citenamefont {Bao},\ and\ \citenamefont
  {Shen}}]{Ji2019}%
  \BibitemOpen
  \bibfield  {author} {\bibinfo {author} {\bibfnamefont {F.}~\bibnamefont
  {Ji}}, \bibinfo {author} {\bibfnamefont {J.}~\bibnamefont {Hu}}, \bibinfo
  {author} {\bibfnamefont {S.}~\bibnamefont {Bao}}, \ and\ \bibinfo {author}
  {\bibfnamefont {H.}~\bibnamefont {Shen}},\ }\href {\doibase
  10.1103/PhysRevC.100.045801} {\bibfield  {journal} {\bibinfo  {journal}
  {Phys. Rev. C}\ }\textbf {\bibinfo {volume} {100}},\ \bibinfo {pages}
  {045801} (\bibinfo {year} {2019})}\BibitemShut {NoStop}%
\bibitem [{\citenamefont {Perot}\ \emph {et~al.}(2020)\citenamefont {Perot},
  \citenamefont {Chamel},\ and\ \citenamefont {Sourie}}]{Perot2020}%
  \BibitemOpen
  \bibfield  {author} {\bibinfo {author} {\bibfnamefont {L.}~\bibnamefont
  {Perot}}, \bibinfo {author} {\bibfnamefont {N.}~\bibnamefont {Chamel}}, \
  and\ \bibinfo {author} {\bibfnamefont {A.}~\bibnamefont {Sourie}},\ }\href
  {\doibase 10.1103/PhysRevC.101.015806} {\bibfield  {journal} {\bibinfo
  {journal} {Phys. Rev. C}\ }\textbf {\bibinfo {volume} {101}},\ \bibinfo
  {pages} {015806} (\bibinfo {year} {2020})}\BibitemShut {NoStop}%
\bibitem [{\citenamefont {Benesty}\ \emph {et~al.}(2009)\citenamefont
  {Benesty}, \citenamefont {Chen}, \citenamefont {Huang},\ and\ \citenamefont
  {Cohen}}]{PC}%
  \BibitemOpen
  \bibfield  {author} {\bibinfo {author} {\bibfnamefont {J.}~\bibnamefont
  {Benesty}}, \bibinfo {author} {\bibfnamefont {J.}~\bibnamefont {Chen}},
  \bibinfo {author} {\bibfnamefont {Y.}~\bibnamefont {Huang}}, \ and\ \bibinfo
  {author} {\bibfnamefont {I.}~\bibnamefont {Cohen}},\ }\href@noop {} {\emph
  {\bibinfo {title} {Pearson Correlation Coefficient. In: Noise Reduction in
  Speech Processing}}},\ Vol.~\bibinfo {volume} {2}\ (\bibinfo  {publisher}
  {Springer},\ \bibinfo {address} {Berlin, Heidelberg},\ \bibinfo {year}
  {2009})\ Chap.\ \bibinfo {chapter} {Pearson Correlation
  Coefficient}\BibitemShut {NoStop}%
\bibitem [{\citenamefont {Abbott}\ \emph {et~al.}(2020)\citenamefont {Abbott}
  \emph {et~al.}}]{Abbott2020}%
  \BibitemOpen
  \bibfield  {author} {\bibinfo {author} {\bibfnamefont {B.~P.}\ \bibnamefont
  {Abbott}} \emph {et~al.} (\bibinfo {collaboration} {LIGO Scientific,
  Virgo}),\ }\href@noop {} {\  (\bibinfo {year} {2020})},\ \Eprint
  {http://arxiv.org/abs/2001.01761} {arXiv:2001.01761 [astro-ph.HE]}
  \BibitemShut {NoStop}%
%%CITATION = ARXIV:2001.01761;%%
\bibitem [{\citenamefont {Tews}\ \emph
  {et~al.}(2018{\natexlab{b}})\citenamefont {Tews}, \citenamefont {Carlson},
  \citenamefont {Gandolfi},\ and\ \citenamefont {Reddy}}]{Tews2018}%
  \BibitemOpen
  \bibfield  {author} {\bibinfo {author} {\bibfnamefont {I.}~\bibnamefont
  {Tews}}, \bibinfo {author} {\bibfnamefont {J.}~\bibnamefont {Carlson}},
  \bibinfo {author} {\bibfnamefont {S.}~\bibnamefont {Gandolfi}}, \ and\
  \bibinfo {author} {\bibfnamefont {S.}~\bibnamefont {Reddy}},\ }\href
  {\doibase 10.3847/1538-4357/aac267} {\bibfield  {journal} {\bibinfo
  {journal} {The Astrophysical Journal}\ }\textbf {\bibinfo {volume} {860}},\
  \bibinfo {pages} {149} (\bibinfo {year} {2018}{\natexlab{b}})}\BibitemShut
  {NoStop}%
\bibitem [{\citenamefont {Maselli}\ \emph {et~al.}(2013)\citenamefont
  {Maselli}, \citenamefont {Cardoso}, \citenamefont {Ferrari}, \citenamefont
  {Gualtieri},\ and\ \citenamefont {Pani}}]{Maselli2013}%
  \BibitemOpen
  \bibfield  {author} {\bibinfo {author} {\bibfnamefont {A.}~\bibnamefont
  {Maselli}}, \bibinfo {author} {\bibfnamefont {V.}~\bibnamefont {Cardoso}},
  \bibinfo {author} {\bibfnamefont {V.}~\bibnamefont {Ferrari}}, \bibinfo
  {author} {\bibfnamefont {L.}~\bibnamefont {Gualtieri}}, \ and\ \bibinfo
  {author} {\bibfnamefont {P.}~\bibnamefont {Pani}},\ }\href {\doibase
  10.1103/PhysRevD.88.023007} {\bibfield  {journal} {\bibinfo  {journal} {Phys.
  Rev. D}\ }\textbf {\bibinfo {volume} {88}},\ \bibinfo {pages} {023007}
  (\bibinfo {year} {2013})}\BibitemShut {NoStop}%
\bibitem [{\citenamefont {Lattimer}()}]{Lattimer2019}%
  \BibitemOpen
  \bibfield  {author} {\bibinfo {author} {\bibfnamefont {J.~M.}\ \bibnamefont
  {Lattimer}},\ }\href@noop {} {}\bibinfo {howpublished} {private
  communication}\BibitemShut {NoStop}%
\bibitem [{\citenamefont {Shane}\ \emph {et~al.}(2015)\citenamefont {Shane},
  \citenamefont {McIntosh}, \citenamefont {Isobe}, \citenamefont {Lynch},
  \citenamefont {Baba}, \citenamefont {Barney}, \citenamefont {Chajecki},
  \citenamefont {Chartier}, \citenamefont {Estee}, \citenamefont {Famiano},
  \citenamefont {Hong}, \citenamefont {Ieki}, \citenamefont {Jhang},
  \citenamefont {Lemmon}, \citenamefont {Lu}, \citenamefont {Murakami},
  \citenamefont {Nakatsuka}, \citenamefont {Nishimura}, \citenamefont {Olsen},
  \citenamefont {Powell}, \citenamefont {Sakurai}, \citenamefont {Taketani},
  \citenamefont {Tangwancharoen}, \citenamefont {Tsang}, \citenamefont
  {Usukura}, \citenamefont {Wang}, \citenamefont {Yennello},\ and\
  \citenamefont {Yurkon}}]{Shane2015}%
  \BibitemOpen
  \bibfield  {author} {\bibinfo {author} {\bibfnamefont {R.}~\bibnamefont
  {Shane}}, \bibinfo {author} {\bibfnamefont {A.~B.}\ \bibnamefont {McIntosh}},
  \bibinfo {author} {\bibfnamefont {T.}~\bibnamefont {Isobe}}, \bibinfo
  {author} {\bibfnamefont {W.~G.}\ \bibnamefont {Lynch}}, \bibinfo {author}
  {\bibfnamefont {H.}~\bibnamefont {Baba}}, \bibinfo {author} {\bibfnamefont
  {J.}~\bibnamefont {Barney}}, \bibinfo {author} {\bibfnamefont
  {Z.}~\bibnamefont {Chajecki}}, \bibinfo {author} {\bibfnamefont
  {M.}~\bibnamefont {Chartier}}, \bibinfo {author} {\bibfnamefont
  {J.}~\bibnamefont {Estee}}, \bibinfo {author} {\bibfnamefont
  {M.}~\bibnamefont {Famiano}}, \bibinfo {author} {\bibfnamefont
  {B.}~\bibnamefont {Hong}}, \bibinfo {author} {\bibfnamefont {K.}~\bibnamefont
  {Ieki}}, \bibinfo {author} {\bibfnamefont {G.}~\bibnamefont {Jhang}},
  \bibinfo {author} {\bibfnamefont {R.}~\bibnamefont {Lemmon}}, \bibinfo
  {author} {\bibfnamefont {F.}~\bibnamefont {Lu}}, \bibinfo {author}
  {\bibfnamefont {T.}~\bibnamefont {Murakami}}, \bibinfo {author}
  {\bibfnamefont {N.}~\bibnamefont {Nakatsuka}}, \bibinfo {author}
  {\bibfnamefont {M.}~\bibnamefont {Nishimura}}, \bibinfo {author}
  {\bibfnamefont {R.}~\bibnamefont {Olsen}}, \bibinfo {author} {\bibfnamefont
  {W.}~\bibnamefont {Powell}}, \bibinfo {author} {\bibfnamefont
  {H.}~\bibnamefont {Sakurai}}, \bibinfo {author} {\bibfnamefont
  {A.}~\bibnamefont {Taketani}}, \bibinfo {author} {\bibfnamefont
  {S.}~\bibnamefont {Tangwancharoen}}, \bibinfo {author} {\bibfnamefont
  {M.~B.}\ \bibnamefont {Tsang}}, \bibinfo {author} {\bibfnamefont
  {T.}~\bibnamefont {Usukura}}, \bibinfo {author} {\bibfnamefont
  {R.}~\bibnamefont {Wang}}, \bibinfo {author} {\bibfnamefont {S.~J.}\
  \bibnamefont {Yennello}}, \ and\ \bibinfo {author} {\bibfnamefont
  {J.}~\bibnamefont {Yurkon}},\ }\href {\doibase
  https://doi.org/10.1016/j.nima.2015.01.026} {\bibfield  {journal} {\bibinfo
  {journal} {Nuclear Instruments and Methods in Physics Research Section A:
  Accelerators, Spectrometers, Detectors and Associated Equipment}\ }\textbf
  {\bibinfo {volume} {784}},\ \bibinfo {pages} {513 } (\bibinfo {year}
  {2015})},\ \bibinfo {note} {{S}ymposium on Radiation Measurements and
  Applications 2014 (SORMA XV)}\BibitemShut {NoStop}%
\bibitem [{\citenamefont {Russotto}\ \emph {et~al.}(2016)\citenamefont
  {Russotto} \emph {et~al.}}]{Russotto2016}%
  \BibitemOpen
  \bibfield  {author} {\bibinfo {author} {\bibfnamefont {P.}~\bibnamefont
  {Russotto}} \emph {et~al.},\ }\href {\doibase 10.1103/PhysRevC.94.034608}
  {\bibfield  {journal} {\bibinfo  {journal} {Phys. Rev. C}\ }\textbf {\bibinfo
  {volume} {94}},\ \bibinfo {pages} {034608} (\bibinfo {year}
  {2016})}\BibitemShut {NoStop}%
\bibitem [{\citenamefont {Fattoyev}\ \emph {et~al.}(2013)\citenamefont
  {Fattoyev}, \citenamefont {Carvajal}, \citenamefont {Newton},\ and\
  \citenamefont {Li}}]{Fattoyev2013}%
  \BibitemOpen
  \bibfield  {author} {\bibinfo {author} {\bibfnamefont {F.~J.}\ \bibnamefont
  {Fattoyev}}, \bibinfo {author} {\bibfnamefont {J.}~\bibnamefont {Carvajal}},
  \bibinfo {author} {\bibfnamefont {W.~G.}\ \bibnamefont {Newton}}, \ and\
  \bibinfo {author} {\bibfnamefont {B.-A.}\ \bibnamefont {Li}},\ }\href
  {\doibase 10.1103/PhysRevC.87.015806} {\bibfield  {journal} {\bibinfo
  {journal} {Phys. Rev. C}\ }\textbf {\bibinfo {volume} {87}},\ \bibinfo
  {pages} {015806} (\bibinfo {year} {2013})}\BibitemShut {NoStop}%
\end{thebibliography}
\end{document}